\documentclass{aa}  

\usepackage{graphicx}
\usepackage{xcolor}
\usepackage{stfloats}
\usepackage{txfonts}
\usepackage{placeins}
\usepackage{float}
\usepackage{lipsum}
\usepackage{subcaption}         
\usepackage{lscape}             
                                
\usepackage[
    colorlinks=true,
    citecolor=blue,
    linkcolor=blue,
    urlcolor=blue
]{hyperref}

\usepackage{booktabs}
\usepackage{tabularx}
 \usepackage{array}

\begin{document}
\raggedbottom
   \title{Photometry-constrained fitting of astrometric microlensing }

\subtitle{Parameter and lens property recovery in Gaia DR4 simulations}

%
%
%

   \author{A. J.  Pucek\inst{\ref{aff:UW}}\corrauth{apucek@astrouw.edu.pl}  
   \and 
   Z. Kaczmarek\inst{\ref{aff:ARI}}\email{zofia.kaczmarek@uni-heidelberg.de}
    \and 
    {\L}. Wyrzykowski\inst{\ref{aff:UW}, \ref{aff:NCBJ},\ref{aff:NOA}}\email{wyrzykow@astrouw.edu.pl}
    \and
    M. Wicker\inst{\ref{aff:UW}}\email{mwicker@astrouw.edu.pl}
    \and
   K. Kruszy\'{n}ska\inst{\ref{aff:LCO}}\email{kkruszyn@andrew.cmu.edu}
   \and
   K. A. Rybicki\inst{\ref{aff:UW}}\email{krybicki@astrouw.edu.pl}
        }

   \institute{
       \label{aff:UW}Astronomical Observatory, University of Warsaw, Al. Ujazdowskie 4, 00-478 Warsaw, Poland 
    \and
    \label{aff:ARI}Astronomisches Rechen-Institut, Zentrum f{\"u}r Astronomie der Universit{\"a}t Heidelberg (ZAH), M{\"o}nchhofstr. 12-14, 69120 Heidelberg, Germany
    \and
    \label{aff:NCBJ}Astrophysics Division, National Centre for Nuclear Research, Pasteura 7, 02-093, Warsaw, Poland
    \and
    \label{aff:NOA}National Observatory of Athens, IAASARS, 
Ioannou Metaxa and Vasileos Pavlou GR-15236 Athens, Greece
\and
    \label{aff:LCO}Las Cumbres Observatory, 6740 Cortona Drive, Suite 102, 93117 Goleta, CA, USA
   }

   \date{}

 
  \abstract
   {Gaia Data Release 4 (DR4) provides epoch photometry and astrometry, opening new opportunities to determine the physical properties of gravitational microlenses.}
   {We test sequential photometric and astrometric modelling for the analysis of the Gaia DR4 microlensing catalogue, assessing the accuracy of angular Einstein radius and lens property recovery, the reliability of formal Einstein-radius uncertainties, and the impact of photometric fitting errors.
} 
   {We perform injection--recovery simulations across diverse microlensing parameters and observing conditions.
With the recovered photometric parameters held fixed,
we fit the angular Einstein radius and the five standard
astrometric parameters assuming a~dark lens.
We examine the effects of blending, baseline magnitude,
and Gaia scanning coverage, and repeat the astrometric
fit using the injected microlensing parameters.}
   {In the absence of blending, recovery accuracy increases with baseline brightness and angular Einstein radius. Among selected configurations, the fraction recovering the angular Einstein radius within 10\% of its injected value increases from 34\% for
$\theta_{\mathrm E}<1$ mas to 87\% for
$\theta_{\mathrm E}\geq6$ mas.
The lens mass and distance are recovered within 20\%
in 65\% and 69\% of selected unblended configurations,
respectively, and both simultaneously in 55\%.
Joint recovery fractions are highest for nearby and massive lenses. 
Blending reduces recovery accuracy and leads to systematic underestimation of the angular Einstein radius, even when the injected photometric parameters are used.
The formal uncertainties fail to fully account for errors arising from blending or photometric fitting. Measurement significance is a useful diagnostic of recovery accuracy without blending, but neither high significance nor photometric membership scores guarantee accurate recovery.
}
   {
Sequential fitting shows promise for analysing large datasets such as Gaia DR4.
The reported recovery fractions depend on the adopted
simulation conditions and do not represent
population-level predictions.
Careful treatment of blending is essential,
as even formally precise measurements may yield
systematically biased parameter estimates.}


   \maketitle
\nolinenumbers

\section{Introduction}

Gravitational microlensing occurs when a~massive foreground
object passes close to the line of sight to a~background
source, producing a~temporary magnification and a~displacement
of the light centroid relative to its unlensed position
\citep{einstein1936,paczynski1986,hog1995,walker1995}.
Since the lens does not need to be luminous, microlensing
provides a~powerful tool for studying otherwise
difficult-to-detect objects, including exoplanets, white
dwarfs, neutron stars, and black holes
\citep{mao1991,bennett2002}.

Originally proposed as a~method for investigating massive
compact halo objects (MACHOs) as potential constituents
of Galactic dark matter \citep{paczynski1986},
microlensing has become the subject of large-scale
photometric surveys, including MACHO, OGLE, EROS,
MOA, and KMTNet
\citep{alcock1992,udalski1992,aubourg1993,bond2001,kim2016}.
However, photometric observations alone generally cannot
determine the lens mass, as the Einstein timescale,
$t_{\rm E}$, depends on both the angular Einstein radius,
$\theta_{\rm E}$, and the relative proper motion,
$\mu_{\rm rel}$ \citep{dominiksahu2000}.
Combined photometric and astrometric observations can
break this degeneracy by constraining $\theta_{\rm E}$
and the microlensing parallax, $\pi_{\rm E}$
\citep{gould2000,dominiksahu2000}.
The potential of this approach has been demonstrated
by the mass measurement of the isolated stellar-mass
black hole OGLE-2011-BLG-0462 using high-precision
Hubble Space Telescope observations
\citep{sahu2022,sahu2025}.

The European Space Agency's \textit{Gaia} mission offers
a~unique opportunity to study microlensing through its
all-sky coverage and combined photometric and astrometric
observations \citep{gaia}. An early example of modelling the expected astrometric
signal for a~\textit{Gaia} microlensing event was presented
by \citet{wyrzykowski2020} for the binary-lens event
Gaia16aye.
Gaia DR3 included 363 microlensing candidate events,
of which 90 were previously unreported
\citep{wyrzykowski2023}.
Furthermore, anomalies in the standard astrometric
solution of GaiaDR3-ULENS-001 have been interpreted
as possible signatures of astrometric microlensing
\citep{jablonska2022}.
However, individual astrometric measurements were not
publicly available in DR3.
\textit{Gaia} Data Release~4%
\footnote{\url{https://www.cosmos.esa.int/web/gaia/release}}
(DR4) provides these measurements,
covering 66 months of observations.

The potential of \textit{Gaia} for astrometric
microlensing studies was recognised well before its
launch \citep{hog1995,belokurov2002}.
Subsequent studies investigated its sensitivity to
astrometric microlensing and its ability to constrain
compact-object masses \citep{rybicki2018}.

More recently, \citet{jankovic2025} developed the
Gaia Astrometric Microlensing Events (GAME) Filter,
demonstrating the possibility of detecting and
characterising events using astrometric observations
alone.  Combined photometric and astrometric modelling has
also been applied to searches for isolated black holes
\citep{lam2022, huston2026}.
The agreement between photometric and astrometric
solutions is particularly important for reliable lens
mass estimates, as demonstrated by separate fits
to both datasets by \citet{lam2022}. Using simulated Galactic populations,
\citet{kaczmarek2026} predicted $322 \pm 70$ events
with potentially detectable astrometric signals
in Gaia DR4, including $209 \pm 57$ events with
detectable photometric signals.
Their joint photometric and astrometric fitting
successfully recovered $t_{\rm E}$ and $\theta_{\rm E}$
for most simulated events, although poorly constrained
microlensing parallaxes limited the accuracy of
lens mass estimates.

In contrast to joint fitting, we investigate a~sequential
approach based on the multi-level photometric pipeline
described by Wyrzykowski et al.\ (in prep.).
We first fit simulated \textit{Gaia} light curves
by minimising $\chi^2$, 
considering blending and annual parallax in separate
photometric models.
The recovered parameters of the parallax model
($t_0$, $u_0$, $t_{\rm E}$, $\pi_{{\rm E},N}$,
and $\pi_{{\rm E},E}$) are then held fixed during
astrometric fitting, allowing $\theta_{\rm E}$ to be
determined alongside the five standard astrometric
parameters through weighted linear least squares.
This approach reduces the complexity of the fit,
which is particularly useful when analysing large
datasets.
The adopted astrometric model assumes a~dark lens
and neglects blending, which may lead to an
underestimation of $\theta_{\rm E}$.

The reliability of this approach depends on both
the adopted models and the available observations.
Sparse sampling and blending can bias the recovered
photometric parameters \citep{wozniak1997}.
The accuracy of $\theta_{\rm E}$ estimates may also
be limited by Gaia's observational cadence,
measurement precision, and one-dimensional along-scan
measurements \citep{lindegren2022}.
Further limitations arise from the partial absorption
of the microlensing signal by the standard
five-parameter astrometric model
\citep{kaczmarek2026b}
and neglected astrometric blending \citep{han1999}.

We assess the performance of sequential fitting
through injection--recovery simulations covering
a~wide range of microlensing parameters and observing
conditions.
We investigate the effects of blending, baseline
magnitude, and scanning coverage on the recovery of
$\theta_{\rm E}$ and the inferred lens masses
and distances.
By repeating the astrometric fit using the injected
microlensing parameters, we assess how much
photometric fitting errors contribute to inaccurate
astrometric solutions. We also assess the reliability of the formal uncertainties in \(\theta_E\) and the usefulness of its measurement significance as a diagnostic of recovery accuracy.
Our aim is to identify the conditions under which
this simplified approach provides reliable parameter
estimates and to assess its potential for analysing
large microlensing datasets such as Gaia DR4.

\section{Simulated Gaia observations}
\label{sec:simulated_observations}

We generated synthetic Gaia photometric and astrometric
observations for three injection--recovery experiments:
an~unblended experiment, a~blending experiment, and a~lens mass--distance experiment. The input parameters
were sampled to explore a~broad range of microlensing
and observing conditions rather than to reproduce
a~Galactic event-rate distribution. 

\subsection{Unblended experiment}
\label{sec:unblended_sampling}

We independently sampled nine base-event parameters from discrete
uniform distributions: the impact parameter $u_0$, the reference epoch of closest approach $t_0$, the Einstein timescale $t_{\mathrm E}$,
the microlensing parallax amplitude $\pi_{\mathrm E}$, its orientation
angle $\phi$, the two components of the source proper motion
$\mu_{\alpha*,\mathrm S}$ and $\mu_{\delta,\mathrm S}$, the source
parallax $\varpi_{\mathrm S}$, and the angular Einstein radius
$\theta_{\mathrm E}$. The adopted ranges and sampling steps are
listed in Table~\ref{tab:injected_parameter_ranges}.

The North and East components of the microlensing parallax were
calculated as
\begin{equation}
  \pi_{\mathrm{E,N}}=\pi_{\mathrm E}\cos\phi,\qquad
  \pi_{\mathrm{E,E}}=\pi_{\mathrm E}\sin\phi.
\end{equation}
The source distance was obtained from
$D_{\mathrm S}/\mathrm{kpc}=(\varpi_{\mathrm S}/\mathrm{mas})^{-1}$.

Following \citet{jankovic2025}, we adopted three sky positions:
$(\alpha,\delta)=(6.5^\circ,-47.3^\circ)$,
$(277.8^\circ,-68.3^\circ)$, and
$(287.9^\circ,45.9^\circ)$. These positions correspond
to 268, 216, and 94 Gaia field-of-view transits,
respectively, over the 66-month DR4 observing period,
allowing us to investigate the effect of scanning
coverage on parameter recovery.

To examine the dependence on baseline brightness, we
considered seven baseline $G$-band magnitudes,
$G_0=13.5,14.5,\ldots,19.5$ mag.

The unblended sample consists of $N_{\mathrm{base}}=1398$
base events: 912 from dedicated unblended simulations
and 486 from the blending experiment
(Sect.~\ref{sec:blending_experiment}), retaining only
configurations with $f_s=1$.
Each base event was simulated across three sky
positions and seven baseline magnitudes, yielding
$3\times7=21$ configurations per base event and
29\,358 simulated configurations in total.
These configurations share the same underlying
microlensing parameters and are therefore not
statistically independent.
This design allows us to investigate the effects
of sky position and baseline brightness on
parameter recovery for the same underlying events.

\begin{table}[t]
\caption{Input parameter sampling in the unblended and
blending experiments. The nine base-event parameters are
sampled independently from discrete uniform distributions.}
\label{tab:injected_parameter_ranges}
\centering
\small
\setlength{\tabcolsep}{2.5pt}
\begin{tabular}{@{}lcll@{}}
\hline\hline
Parameter & Range/values & Step & Sampling \\
\hline
$u_0$
  & $[-1.5,1.5]$ & $0.05$ & Discrete uniform \\
$t_0$ [yr]
  & $[2014.5,2020.0]$ & $0.01$ & Discrete uniform \\
$t_{\mathrm E}$ [d]
  & $[20,500]$ & $1$ & Discrete uniform \\
$\pi_{\mathrm E}$
  & $[0.01,1.00]$ & $0.01$ & Discrete uniform \\
$\phi$ [deg]
  & $[0,359]$ & $1$ & Discrete uniform \\
$\mu_{\alpha*,\mathrm S}$ [mas yr$^{-1}$]
  & $[-10,10]$ & $0.1$ & Discrete uniform \\
$\mu_{\delta,\mathrm S}$ [mas yr$^{-1}$]
  & $[-10,10]$ & $0.1$ & Discrete uniform \\
$\varpi_{\mathrm S}$ [mas]
  & $[0.1,1.0]$ & $0.01$ & Discrete uniform \\
$\theta_{\mathrm E}$ [mas]
  & $[0.05,10.00]$ & $0.05$ & Discrete uniform \\
\hline
$G_0$ [mag]
  & $13.5,\ldots,19.5$ & $1.0$ & Fixed grid \\
$f_s$ (unblended)
  & $1.0$ & -- & Fixed \\
$f_s$ (blended)
  & $0.1,\ldots,0.9,0.95,1.0$ & -- & Fixed grid \\
\hline
$(\alpha_1,\delta_1)$ [deg]
  & $(6.5,-47.3)$ & -- & Fixed \\
$(\alpha_2,\delta_2)$ [deg]
  & $(277.8,-68.3)$ & -- & Fixed \\
$(\alpha_3,\delta_3)$ [deg]
  & $(287.9,45.9)$ & -- & Fixed \\
\hline
\end{tabular}
\end{table}

\subsection{Blending experiment}
\label{sec:blending_experiment}

To investigate the effect of blending on parameter recovery,
we repeated the experiment with variable source-flux fractions,
defined as
\begin{equation}
    f_s = \frac{F_{\mathrm S}}{F_{\mathrm S}+F_{\mathrm B}},
\end{equation}
where $F_{\mathrm S}$ is the unlensed source flux and
$F_{\mathrm B}$ is the lens flux.
The baseline magnitude $G_0$ corresponds to their
combined unlensed flux.

The sample consists of $N_{\mathrm{base,bl}}=486$ base events
generated using the parameter ranges listed in
Table~\ref{tab:injected_parameter_ranges}.
Each base event was simulated for three sky positions,
seven baseline magnitudes, and eleven source-flux fractions,
$f_s\in\{0.1,0.2,\ldots,0.9,0.95,1.0\}$.
The resulting sample comprises
$N_{\mathrm{blend}}=231N_{\mathrm{base,bl}}=112\,266$
configurations.
The value $f_s=0.95$ probes the weak-blending regime,
while $f_s=1$ provides an unblended reference.

Again, the resulting configurations share
the same underlying microlensing parameters and
are not statistically independent.
This design enables controlled comparisons of
the effects of blending on parameter recovery
under different observing conditions.

\subsection{Lens mass--distance experiment}
\label{sec:lens_mass_distance}

To investigate parameter recovery as a function of the
physical properties of the lens, we constructed a stratified
sample in the $(M_{\mathrm L},D_{\mathrm L})$ plane,
adopting the ranges explored by \citet{jankovic2025}.

The lens-mass interval $[0.1,20]\,M_\odot$ and the
lens-distance interval $[0.1,7.9]$ kpc were divided into
20 logarithmically and eight linearly spaced bins,
respectively. We generated $N_{\mathrm{base,MD}}=40\,000$
base events distributed approximately uniformly across
the resulting 160 cells. Within each cell, lens masses
were sampled uniformly in $\log M_{\mathrm L}$ and
distances uniformly in $D_{\mathrm L}$.

For each lens, the source parallax $\varpi_{\mathrm S}$
was sampled from a discrete uniform distribution,
subject to $D_{\mathrm S}>D_{\mathrm L}$ and the
injection constraints listed in
Table~\ref{tab:lens_mass_distance_sampling}.
The relative parallax, angular Einstein radius,
and microlensing parallax were calculated as
\begin{align}
    \pi_{\mathrm{rel}} &= \varpi_{\mathrm L}
                         -\varpi_{\mathrm S},\\
    \theta_{\mathrm E} &=
    \sqrt{\kappa M_{\mathrm L}\pi_{\mathrm{rel}}},\\
    \pi_{\mathrm E} &= \frac{\pi_{\mathrm{rel}}}
                           {\theta_{\mathrm E}},
\end{align}
where $\varpi_{\mathrm L}/\mathrm{mas}
=(D_{\mathrm L}/\mathrm{kpc})^{-1}$ and
$\kappa=8.144\,\mathrm{mas}\,M_\odot^{-1}$.

Unlike in the preceding experiments, $\theta_{\mathrm E}$
and $\pi_{\mathrm E}$ were derived from the lens mass
and geometry rather than sampled independently.
The remaining parameters were sampled as specified
in Table~\ref{tab:lens_mass_distance_sampling},
without imposing a Galactic velocity distribution.

We adopted a single sky position,
$(\alpha,\delta)=(277.8^\circ,-68.3^\circ)$,
to keep the Gaia scanning geometry fixed.

Of the 40\,000 base events, 20\,000 were simulated
both with $f_s=1$ and with a variable source-flux
fraction, while the remaining 20\,000 were simulated
only with a variable source-flux fraction.
The resulting sample comprises 60\,000 configurations:
20\,000 in the unblended variant and 40\,000 in the
blended variant.

The baseline magnitudes and source-flux fractions
were distributed approximately evenly across their
respective grids within each mass--distance cell.

\begin{table}[t]
\caption{Input parameter sampling in the lens mass--distance
experiment. Lens masses and distances are stratified across
160 cells. The blended variant includes $f_s=1$.}
\label{tab:lens_mass_distance_sampling}
\centering
\small
\setlength{\tabcolsep}{2.5pt}
\begin{tabular}{@{}lcll@{}}
\hline\hline
Parameter & Range/values & Step & Sampling \\
\hline
$M_{\mathrm L}$ [$M_\odot$]
  & $[0.1,20]$ & -- & Log bins (20) \\
$D_{\mathrm L}$ [kpc]
  & $[0.1,7.9]$ & -- & Linear bins (8) \\
$\varpi_{\mathrm S}$ [mas]
  & $[0.01,2.00]$ & $0.01$ & Conditional \\
$\theta_{\mathrm E}$ [mas]
  & $\geq0.05$ & -- & Derived \\
$\pi_{\mathrm E}$
  & $\geq0.01$ & -- & Derived \\
$u_0$
  & $[-1.5,1.5]$ & $0.05$ & Discrete uniform \\
$t_0$ [yr]
  & $[2014.5,2020.0]$ & $0.01$ & Discrete uniform \\
$t_{\mathrm E}$ [d]
  & $[20,500]$ & $1$ & Discrete uniform \\
$\phi$ [deg]
  & $[0,359]$ & $1$ & Discrete uniform \\
$\mu_{\alpha*,\mathrm S}$ [mas yr$^{-1}$]
  & $[-10,10]$ & $0.1$ & Discrete uniform \\
$\mu_{\delta,\mathrm S}$ [mas yr$^{-1}$]
  & $[-10,10]$ & $0.1$ & Discrete uniform \\
$G_0$ [mag]
  & $13.5,\ldots,19.5$ & $1.0$ & Balanced \\
$f_s$ (unblended)
  & $1.0$ & -- & Fixed \\
$f_s$ (blended)
  & $0.1,\ldots,0.9,0.95,1.0$ & -- & Balanced \\
$(\alpha,\delta)$ [deg]
  & $(277.8,-68.3)$ & -- & Fixed \\
\hline
\end{tabular}

\end{table}

\subsection{Generation of simulated Gaia observations}
\label{sec:mock_generation}

Synthetic photometric and astrometric observations were
generated using the \texttt{astromet} package\footnote{
\url{https://github.com/zpenoyre/astromet.py}} \citep{penoyre2022}.
Observation epochs and scanning angles were obtained from
\texttt{gaiascanlaw}\footnote{
\url{https://github.com/zpenoyre/gaiascanlaw}}
for the 66-month Gaia DR4 observing period.

For each configuration, \texttt{astromet.track} was used
to compute the apparent astrometric trajectory and
microlensing-induced magnitude variations, accounting for
source proper motion, annual parallax, and blending
due to lens light.

The magnitude-dependent photometric uncertainty per
field-of-view transit, $\sigma_{\mathrm{tr}}$, was
modelled using the third-order polynomial adopted by
\citet{kaczmarek2026}~from~\citet{eyer2015}: 

\begin{equation}
\log_{10}\sigma_{\mathrm{tr}}(G)\simeq
0.001404G^3-0.0563G^2+0.958G-9.249,
\label{eq:photometric_uncertainty}
\end{equation}
where $G$ is the instantaneous magnitude and
$\sigma_{\mathrm{tr}}$ is expressed in magnitudes.

Each transit was represented by nine independent CCD
magnitude measurements with Gaussian noise and
$\sigma_{\mathrm{CCD}}=3\sigma_{\mathrm{tr}}$.
The arithmetic mean of the nine CCD magnitudes was
used as a single photometric measurement per transit.

Along-scan astrometric observations were generated
using \texttt{astromet.mock\_obs}, also with nine
CCD measurements per transit and magnitude-dependent
uncertainties provided by \texttt{astromet.sigma\_ast}.
A standard five-parameter astrometric solution was
also obtained for each configuration, adopting the
reference epoch 2017.5.

In the blending experiment, the same standardised noise
realisations were used across all values of $f_s$ for
a given base event, sky position, and baseline magnitude.
This allows for a controlled comparison of the effect
of blending on parameter recovery.

\section{Photometric fitting}
\label{sec:photometric_fitting}

The photometric fitting procedure follows the Gaia DR4
pipeline (Wyrzykowski et al., in prep.), which builds on
the three-level microlensing modelling scheme described
by \citet{wyrzykowski2023}.
We implemented this procedure in Python, retaining
the original model hierarchy, model-selection criteria,
and membership-scoring approach.
Our implementation was adapted to simulated $G$-band
observations and applied consistently across all
three experiments.
Where our implementation differs from the original pipeline, these differences are noted in the following sections.

\subsection{Photometric models}
\label{sec:photometric_models}

The point-source, point-lens magnification is given by
\begin{equation}
A(u)=\frac{u^2+2}{u\sqrt{u^2+4}},
\qquad
u(t)=\sqrt{u_0^2+
\left(\frac{t-t_0}{t_{\mathrm E}}\right)^2},
\label{eq:paczynski_magnification}
\end{equation}
where $u_0$ is the impact parameter, $t_0$ is the
epoch of closest approach in the absence of parallax,
and $t_{\mathrm E}$ is the Einstein timescale.
The expression for $u(t)$ assumes rectilinear
relative motion.

Level~0 fits the standard Paczyński model with
four free parameters $(u_0,G_0,t_0,t_{\mathrm E})$,
with the modelled magnitude given by
$G(t)=G_0-2.5\log_{10}A(u(t))$.

Since blending and annual microlensing parallax
can produce partially degenerate distortions of
the light curve, they are considered separately
in Levels~1 and~2.

Level~1 introduces the source-flux fraction
$f_s$ as a~fifth parameter:
\begin{equation}
G(t)=G_0-2.5\log_{10}
\left[1+f_s(A(u(t))-1)\right],
\label{eq:photometric_model}
\end{equation}
with $0\leq f_s\leq1$.

Level~2 accounts for annual microlensing parallax
instead of blending, introducing two additional
parameters, $\pi_{\mathrm{E,N}}$ and
$\pi_{\mathrm{E,E}}$.
The source--lens separation vector is given by
\begin{equation}
\boldsymbol{u}(t)=
\boldsymbol{u}_{\mathrm{lin}}(t)
+\boldsymbol{u}_{\mathrm{par}}(t),
\label{eq:parallax_trajectory}
\end{equation}
where the parallax term depends on the microlensing
parallax vector and the observer's parallax factors.
The magnification is evaluated at
$u(t)=|\boldsymbol{u}(t)|$, assuming $f_s=1$.

The same three photometric models were fitted
in all three experiments, regardless of whether
blending was included in the simulations.

\subsection{Optimisation and parameter uncertainties}
\label{sec:photometric_optimisation}

The Gaia DR4 photometric pipeline uses observations
in the $G$, $G_{\rm BP}$, and $G_{\rm RP}$ bands,
whereas our simulations include only $G$-band
photometry.
To account for this difference, we adopted an
empirical scaling of the photometric uncertainties
during initial event selection and photometric fitting:

\begin{equation}
\sigma_{\mathrm{fit},i}
= \sqrt{10}\,\sigma_{\mathrm{tr}}(G_i),
\label{eq:photometric_error_scaling}
\end{equation}

where $\sigma_{\mathrm{tr}}$ is defined in
Eq.~\ref{eq:photometric_uncertainty}.
The simulated observational noise was not modified.
For fixed residuals, this scaling reduces the
photometric $\chi^2$ by a factor of ten.

Before fitting, we required at least three
observations to be brighter than the median
magnitude by more than $3\sigma_{\mathrm{fit},i}$.

All models were fitted by minimising $\chi^2$
using the Nelder--Mead simplex algorithm implemented
in \texttt{SciPy}, with photometric uncertainties
$\sigma_{\mathrm{fit},i}$ defined in
Eq.~\ref{eq:photometric_error_scaling}.
The maximum numbers of function evaluations
per attempt were 15\,000, 5\,000, and 2\,000
for Levels~0, 1, and~2, respectively. Although the same optimisation algorithm is used
in the original Gaia pipeline, its implementation
and stopping conditions may differ from those
adopted here.

Initial Level~0 parameters were estimated from
the observed light-curve amplitude, the 90th
percentile of the observed magnitudes, and the
epoch of maximum observed brightness, adopting
an initial timescale of $t_{\mathrm E}=60$ days.

Levels~1 and~2 were initialised independently
from the Level~0 solution, adopting $f_s=1$
and $\pi_{\mathrm{E,N}}=\pi_{\mathrm{E,E}}=0$,
respectively.
Failure to converge at Level~0 or Level~1
terminated the fitting procedure.

For Level~2, the positive and negative $u_0$
branches were optimised separately, with
25 attempts per branch, including 24
randomised starting points.
The initial values of $|u_0|$, $t_0$,
$t_{\mathrm E}$, and both microlensing parallax
components were varied, while the initial $G_0$
was taken from Level~0.
The converged solution with the lowest $\chi^2$
was retained for each branch.

Parameter uncertainties were estimated from
one-dimensional sections of the $\chi^2$ surface,
holding all other parameters fixed.
We adopted $\Delta\chi^2$ thresholds of
4.72, 5.89, and 7.04 for Levels~0, 1, and~2,
respectively.
The uncertainty was defined as half the distance
between the two threshold crossings, when available.
These estimates neglect parameter correlations
and do not represent marginalised $1\sigma$
uncertainties.

\subsection{Model selection and membership score}
\label{sec:model_selection}

The best-fitting photometric model was selected
as the converged solution with the lowest
reduced $\chi^2$.

A~penalty-based membership score was calculated
separately for each converged Level~2 branch.
The score accounts for the light-curve amplitude,
goodness of fit, and fitted parameter values
and uncertainties.
A~score of at least 0.8 was adopted as the
membership criterion.
The detailed definition of the score and its
settings are described in the Gaia DR4 documentation.

Regardless of which photometric model provided
the best fit, both converged Level~2 branches
were retained for astrometric analysis, when available.

\begin{figure}[!htbp]

    \centering
    \includegraphics[
    width=\columnwidth,
    height=0.78\textheight,
    keepaspectratio
]{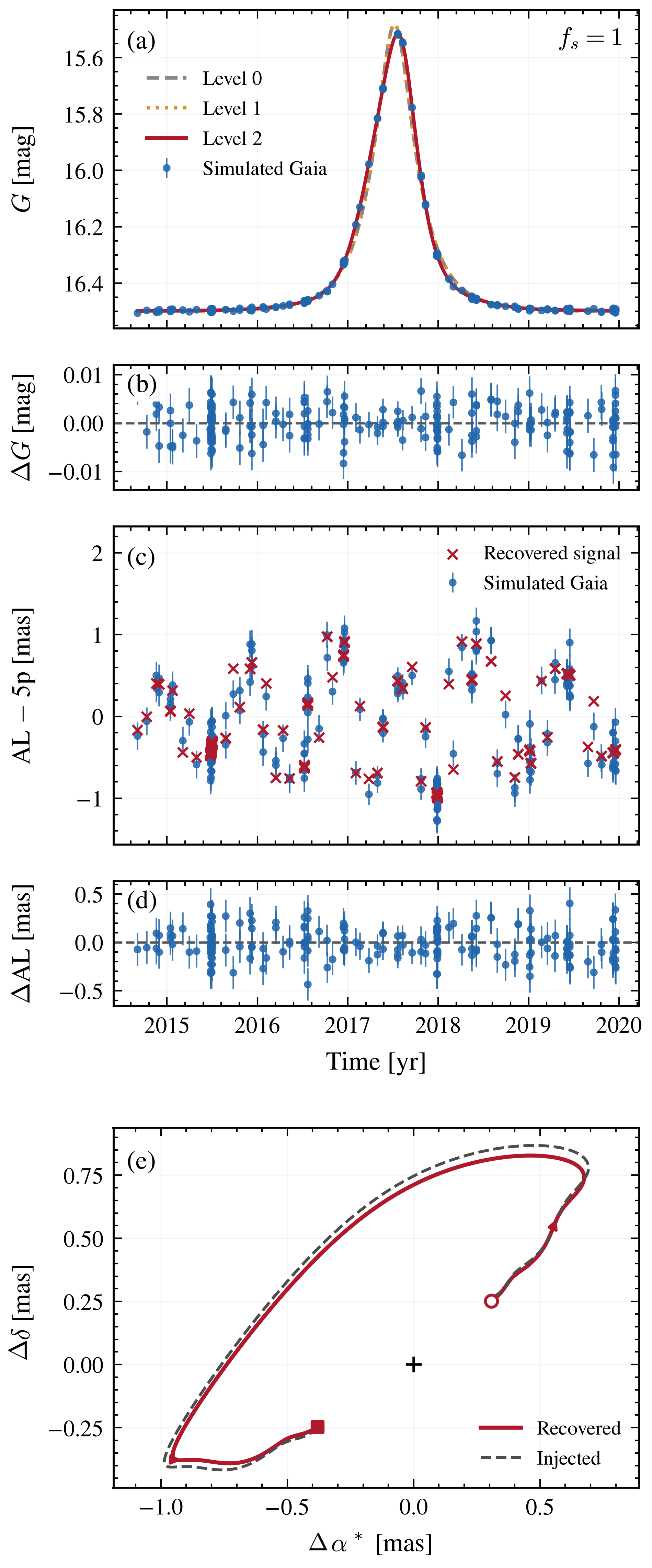}
\caption{
    Illustrative example of sequential photometric
    and astrometric fitting of a simulated
    microlensing event without blending ($f_s=1$).
    (a) Simulated Gaia photometric observations
    and the three fitted photometric models.
    (b) Photometric residuals relative to the
    selected Level~2 solution.
    (c) Along-scan astrometric observations after
    subtracting the fitted five-parameter model,
    together with the recovered microlensing signal.
    (d) Residuals of the astrometric fit.
    (e) Injected and recovered astrometric
    microlensing trajectories in the plane of
    the sky. The open circle and filled square
    mark the beginning and end of the recovered
    trajectory, respectively. The arrows indicate
    the direction of motion, and 
the black cross indicates zero
astrometric displacement relative
to the unlensed source trajectory. 
For visualisation, the astrometric observations
are averaged per field-of-view transit, while
the fitting uses individual CCD measurements. }
\label{fig:example_microlensing_fit}
\end{figure}
 
\section{Astrometric modelling}
\label{sec:astrometric_modelling}

To determine the angular Einstein radius
$\theta_{\mathrm E}$, we modelled the astrometric
observations using the microlensing parameters
recovered from the photometric analysis
(Sect.~\ref{sec:photometric_fitting}).
By holding these parameters fixed, we reduced
the astrometric fitting to a~linear
least-squares problem.

\subsection{Astrometric microlensing model}
\label{sec:astrometric_model}

The unlensed source motion was described by the
standard five-parameter astrometric model
\citep{lindegren2012}, comprising two position
offsets at the reference epoch,
$\Delta\alpha_0^*$ and $\Delta\delta_0$,
the source parallax $\varpi_{\mathrm S}$,
and two proper-motion components,
$\mu_{\alpha*,\mathrm S}$ and
$\mu_{\delta,\mathrm S}$.
The asterisk denotes multiplication of the
right-ascension component by $\cos\delta$.

For a~point source lensed by a~dark point-mass
lens, the displacement of the unresolved image
centroid relative to the unlensed source is
\citep{hog1995,walker1995,miyamoto1995}
\begin{equation}
\boldsymbol{\delta\theta}(t)=
\theta_{\mathrm E}
\frac{\boldsymbol{u}(t)}
{|\boldsymbol{u}(t)|^2+2},
\label{eq:centroid_displacement}
\end{equation}
where $\boldsymbol{u}(t)$ is the position of
the source relative to the lens in units of
$\theta_{\mathrm E}$, including annual
microlensing parallax.

Gaia epoch astrometry is expressed in local
plane coordinates (LPCs; \citealt{lindegren2022}).
The predicted along-scan (AL) position is 

\begin{equation}
w_{\mathrm{model},i}
=w_{5p,i}
+\theta_{\mathrm E}
\frac{\boldsymbol{u}(t_i)
\cdot\boldsymbol{e}_{w,i}}
{|\boldsymbol{u}(t_i)|^2+2}
\equiv w_{5p,i}+\theta_{\mathrm E}h_i,
\label{eq:astrometric_microlensing}
\end{equation}
where $w_{5p,i}$ is the AL position predicted
by the five-parameter model and $h_i$ is the
dimensionless projected microlensing displacement.
The AL unit vector is
$\boldsymbol{e}_{w,i}=(\cos\psi_i,\sin\psi_i)$
in North--East coordinates, with the scanning
angle $\psi_i$ measured from North towards East. The astrometric fitting model assumes a~dark
lens ($f_s=1$), irrespective of the injected
source-flux fraction. 

\subsection{Linear astrometric fitting}
\label{sec:linear_astrometric_fitting}

For each converged photometric Level~2 branch,
the recovered parameters
$(u_0,t_0,t_{\mathrm E},
\pi_{\mathrm{E,N}},\pi_{\mathrm{E,E}})$
were held fixed.
The resulting microlensing trajectory determines
$h_i$ at each observation epoch, leaving
$\theta_{\mathrm E}$ as a~linear scaling
parameter.

The five standard astrometric parameters and
$\theta_{\mathrm E}$ were fitted simultaneously
using weighted linear least squares, minimising
the astrometric $\chi^2_{\mathrm{AL}}$ and taking
into account the individual AL measurement
uncertainties.

Formal parameter uncertainties were obtained
as the square roots of the diagonal elements
of the covariance matrix.
These uncertainties are conditional on the
fixed photometric parameters and do not
propagate their estimation errors.
No additional astrometric excess noise
was fitted.

\subsection{Solution selection}
\label{sec:astrometric_solution_selection}

The available Level~2 branches were fitted
independently.
The angular Einstein radius was treated as an
unconstrained linear coefficient, allowing
negative fitted values.

If two successful solutions yielded Einstein
radii of opposite signs, the solution with
positive $\theta_{\mathrm E}$ was preferred.
Otherwise, the solution with the lower
astrometric $\chi^2_{\mathrm{AL}}$ was selected.
If only one branch was successfully fitted,
it was retained irrespective of the sign
of $\theta_{\mathrm E}$.

The selected astrometric solution inherited
the membership score of its corresponding
photometric Level~2 branch.
Solutions with scores $\geq0.8$ were flagged
as passing the membership criterion.

An illustrative example of the sequential
photometric and astrometric fitting procedure
for an unblended microlensing event is presented
in Fig.~\ref{fig:example_microlensing_fit}.

\subsection{Oracle experiment}
\label{sec:oracle_experiment}

To isolate the impact of photometric parameter
recovery errors on the astrometric solution,
we repeated the astrometric fitting for the
unblended and blending experiments using the
injected rather than the recovered microlensing
parameters. The same six astrometric parameters were fitted
to the same simulated observations, preserving
their noise realisations and measurement
uncertainties.

The oracle experiment allows us to assess
the recovery of $\theta_{\mathrm E}$ independently
of photometric fitting errors.
The results remain  affected by measurement noise,
Gaia scanning coverage, and the presence of blending.

\section{Recovery criteria}
\label{sec:recovery_criteria}

We evaluated the pipeline performance by comparing
the recovered parameters with their injected values.
We distinguished between successful fits, fitting
failures, and configurations rejected by the
initial photometric selection criterion.
Successful fitting does not necessarily imply
accurate parameter recovery.

We defined two recovery criteria, P10 and P20,
corresponding to relative errors below 10\%
and 20\%, respectively:
\begin{equation}
\frac{|p_{\mathrm{rec}}-p_{\mathrm{inj}}|}
{|p_{\mathrm{inj}}|}<\epsilon,
\qquad \epsilon=0.1,\ 0.2,
\label{eq:recovery_criterion}
\end{equation}
where $p_{\mathrm{inj}}$ and $p_{\mathrm{rec}}$
denote the injected and recovered parameter values,
respectively.
These criteria were applied to the angular Einstein
radius $\theta_{\mathrm E}$, the lens mass
$M_{\mathrm L}$, and the lens distance
$D_{\mathrm L}$.

Since the lens mass and distance were not fitted
directly, they were calculated from the recovered
angular Einstein radius $\theta_{\mathrm E}$,
microlensing parallax $\pi_{\mathrm E}$,
and source parallax $\varpi_{\mathrm S}$:
\begin{equation}
M_{\mathrm{L,rec}}=
\frac{\theta_{\mathrm{E,rec}}}
{\kappa\pi_{\mathrm{E,rec}}},
\qquad
\varpi_{\mathrm{L,rec}}=
\varpi_{\mathrm{S,rec}}+
\theta_{\mathrm{E,rec}}\pi_{\mathrm{E,rec}},
\label{eq:recovered_lens_properties}
\end{equation}
where $\pi_{\mathrm{E,rec}}$ is the magnitude
of the recovered microlensing parallax vector.
The lens distance was obtained from
$D_{\mathrm{L,rec}}/\mathrm{kpc}
=(\varpi_{\mathrm{L,rec}}/\mathrm{mas})^{-1}$.
Joint P10 and P20 recovery required both
$M_{\mathrm L}$ and $D_{\mathrm L}$ to satisfy
the corresponding recovery criterion.

Configurations with successful astrometric fits
and membership scores $\geq 0.8$ are hereafter
referred to as selected configurations. Recovery fractions were calculated relative
to either all simulated configurations or
only the selected configurations, as specified
in each case.

Importantly, accurate parameter recovery was defined by
the P10 and P20 criteria, based solely on the
relative differences between recovered and
injected parameter values, without imposing
additional physical-validity constraints. The occurrence of non-positive parameter estimates is investigated separately in
Sect.~\ref{sec:MD_recovery_results}.

To assess the formal significance of the recovered angular
Einstein radius and the reliability of its estimated uncertainty,
we define
\begin{equation}
    S_{\theta_{\rm E}} =
    \frac{\theta_{\rm E,rec}}{\sigma_{\theta_{\rm E}}},
    \qquad
    z_{\theta_{\rm E}} =
    \frac{\theta_{\rm E,rec}-\theta_{\rm E,inj}}
         {\sigma_{\theta_{\rm E}}},
\end{equation}
where $\sigma_{\theta_{\rm E}}$ is the formal uncertainty
obtained from the astrometric fit (Sect.~\ref{sec:linear_astrometric_fitting}).
While $S_{\theta_{\rm E}}$ quantifies the formal significance
of the measurement, the standardised residual
$z_{\theta_{\rm E}}$ allows us to assess the empirical
coverage of the estimated uncertainties.
For unbiased Gaussian errors with correctly estimated
uncertainties, approximately 68.3\%, 95.4\%, and 99.7\%
of configurations are expected to satisfy
$|z_{\theta_{\rm E}}|\leq 1$, 2, and 3, respectively.

\section{Results}

\subsection{Pipeline performance without blending}
\label{sec:results_unblended}

We evaluated the pipeline using 29\,358
unblended configurations ($f_s=1$), based on
1\,398 underlying events simulated at three
sky positions and seven baseline magnitudes.
These configurations are therefore not
statistically independent.

Photometric fitting succeeded for 69.15\% of
configurations and failed for 28.47\%, while
2.38\% were rejected during initial selection.
Nearly all fitting failures (99.96\%) occurred at Level~1, where the optimiser reached the maximum number of function evaluations. The reported failure fraction is therefore dependent on the adopted optimisation settings. All successful photometric fits also yielded successful astrometric solutions.

Table~\ref{tab:unblended_recovery} summarises
the recovery results. Among all injected
configurations, 46.66\% satisfied P10 and
53.04\% satisfied P20. The membership criterion selected 50.57\%
of all configurations. Among these, 75.75\%
satisfied P10 and 85.31\% satisfied P20.
However, among successfully fitted configurations
that did not pass the membership criterion,
44.97\% satisfied P10 and 53.27\% satisfied P20.

Figure~\ref{fig:member_recovery} compares the injected
and recovered photometric and astrometric parameters.
While the baseline magnitude, reference epoch, and source
proper motions are generally well recovered, larger
scatter is observed for the Einstein timescale,
microlensing parallax components, angular Einstein
radius, and source parallax.

\begin{table}[t]
\centering
\caption{Pipeline performance and angular Einstein-radius
recovery in the unblended sample ($f_s=1$).}
\label{tab:unblended_recovery}
\small
\setlength{\tabcolsep}{3pt}
\renewcommand{\arraystretch}{1.2}

\begin{tabular}{lrrrr}
\hline\hline
Sample & $N$ & Fraction & P10 & P20 \\
       &     & [\%] & [\%] & [\%] \\
\hline
All injected
 & 29358 & 100.00 & 46.66 & 53.04 \\
Passed pipeline
 & 20300 & 69.15 & 67.48 & 76.70 \\
\hline
\textbf{Selected}
 & \textbf{14847}
 & \textbf{50.57}
 & \textbf{75.75}
 & \textbf{85.31} \\
Not selected
 & 5453 & 18.57 & 44.97 & 53.27 \\
\hline
\end{tabular}

\tablefoot{
Fraction denotes the percentage of all injected
configurations.
P10 and P20 are calculated relative to the number
of configurations in each row.
The selected sample includes successful astrometric
fits with membership scores $\geq0.8$, while the
not-selected sample includes successful astrometric
fits with scores below 0.8.
}
\end{table}

\subsection{Dependence on Gaia scanning coverage}
\label{sec:scanning_coverage}

The selection fractions are similar across the three
investigated sky positions, ranging from 50.07\%
to 51.52\%. Despite substantial differences in the
number of Gaia transits (94--268), the P10 recovery
fractions among selected configurations range from
73.19\% to 77.09\%, while the P20 fractions range
from 83.36\% to 86.30\%.
The detailed results are presented in
Table~\ref{tab:sky_comparison} in
Appendix~\ref{app:additional_results}.

Given the relatively small differences between the
three sky positions, we combine them in the following
analyses. The lens mass--distance experiment uses
only Field 2.

\subsection{Effect of blending}
\label{sec:blending_recovery}

We investigated the effect of blending on
$\theta_{\mathrm E}$ recovery using 112\,266
configurations, with 10\,206 for each of the
eleven source-flux fractions. Since these
configurations share the same underlying events,
they are not statistically independent.

The detailed selection and recovery fractions
for each value of $f_s$ are presented in
Table~\ref{tab:blending_recovery}
in Appendix~\ref{app:blending_recovery}.

As blending increases from $f_s=1$ to $f_s=0.1$,
the selection fraction decreases from 50.96\%
to 11.12\%.
Among the selected configurations, the P10
recovery fraction decreases from 77.16\% at
$f_s=1$ to 37.86\% at $f_s=0.7$,
4.59\% at $f_s=0.5$, and 0.62\% at $f_s=0.1$.
The P20 recovery fractions show a similar trend.

The residual distributions of the recovered
photometric and astrometric parameters for
different source-flux fractions are presented
in Appendix~\ref{app:residuals}.

\subsection{Factors affecting Einstein-radius recovery}
\label{sec:magnitude_recovery}

\begin{figure}[t]
    \centering
    \includegraphics[width=0.9\columnwidth]
    {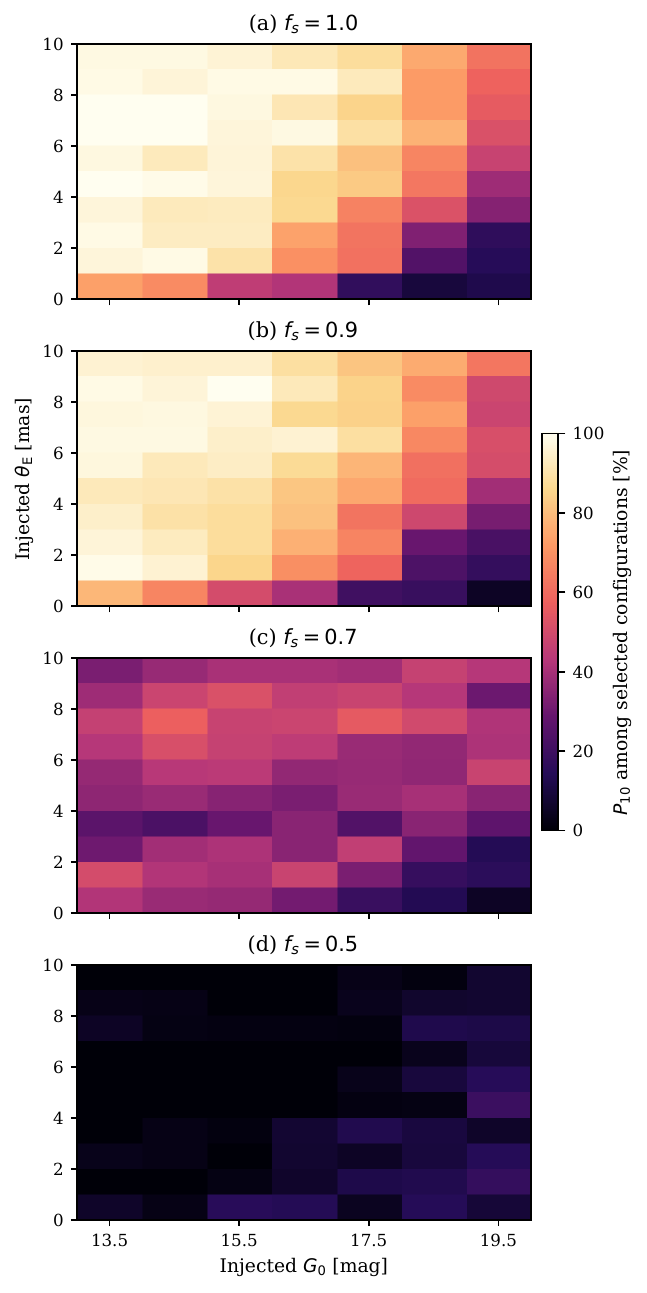}

    \caption{
    P10 recovery fraction among selected
    configurations as a function of the injected
    baseline magnitude $G_0$ and angular Einstein
    radius $\theta_{\mathrm E}$ for four
    source-flux fractions:
    $f_s=0.5$, $0.7$, $0.9$, and $1.0$.
    All panels share the same colour scale.
    }
    \label{fig:blending_grid_p10}
\end{figure}

Figure~\ref{fig:blending_grid_p10} shows the
P10 recovery fraction among selected configurations
as a function of the injected baseline magnitude
$G_0$ and angular Einstein radius $\theta_{\mathrm E}$
for four source-flux fractions.

At $f_s=0.5$, accurate recovery is rare
throughout the explored parameter space.
At $f_s=0.7$, the recovery fraction increases,
although no clear systematic dependence on
$G_0$ or $\theta_{\mathrm E}$ is observed.
A clearer dependence on both parameters emerges
at $f_s=0.9$ and $f_s=1.0$, with higher recovery
fractions at brighter baseline magnitudes
and larger Einstein radii.

Among selected unblended configurations,
the P10 recovery fraction increases from
33.52\% for $\theta_{\rm E}<1$ mas to
86.72\% for $\theta_{\rm E}\geq6$ mas.
The corresponding P20 fractions are 49.11\%
and 92.58\%, respectively.

Recovery accuracy also increases with the formal
significance of the angular Einstein-radius measurement.
For $0\leq S_{\theta_{\rm E}}<3$, the P10 and P20
fractions are 11.52\% and 23.52\%, respectively,
increasing to 91.38\% and 96.81\% for
$S_{\theta_{\rm E}}\geq10$.
Detailed results for the unblended sample
are presented in Table~\ref{tab:theta_snr_recovery}.

In the presence of blending, however, high formal
significance does not necessarily correspond to
accurate recovery.
At $f_s=0.5$, 66.23\% of selected configurations
satisfy $S_{\theta_{\rm E}}\geq10$, yet only
2.37\% of these satisfy P10 and 26.78\% satisfy P20
(Table~\ref{tab:blending_uncertainties}).

\subsection{Impact of photometric fitting errors
on astrometric recovery}
\label{sec:oracle_recovery}

To investigate the impact of photometric fitting
errors on the recovery of $\theta_{\mathrm E}$,
we compare the standard pipeline with the oracle
approach, in which the injected photometric
parameters are held fixed during astrometric fitting.

Both approaches are evaluated on the same
configurations, selected by the standard pipeline.
As reported in Tables~\ref{tab:unblended_recovery}
and~\ref{tab:blending_recovery}, these comprise
14\,847 configurations from the unblended experiment
and 42\,965 from the blending experiment.

For the unblended experiment, the oracle increases
the P10 recovery fraction from 75.75\% to 81.61\%
and the P20 fraction from 85.31\% to 90.46\%.
For the blending experiment, the corresponding
fractions increase from 41.27\% to 46.02\%
and from 57.81\% to 62.90\%, respectively.

\begin{figure*}[!t]    \centering
    \includegraphics[width=0.85\textwidth]
    {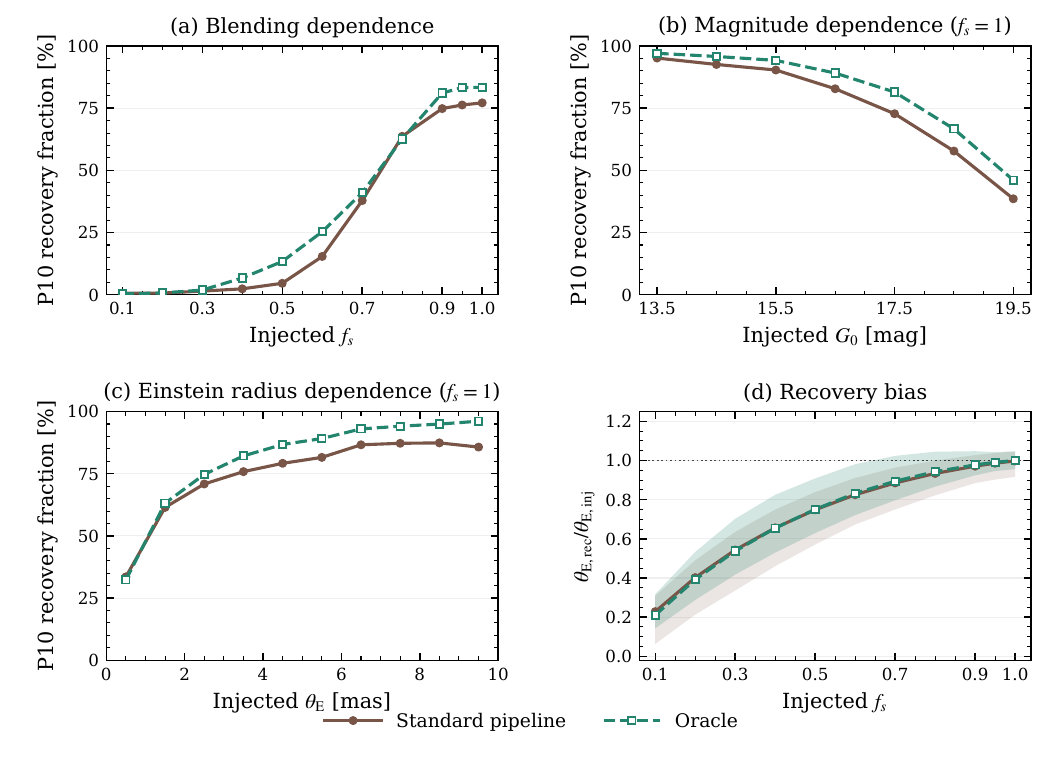}

    \caption{
    Comparison of angular Einstein-radius recovery
    using the standard pipeline (solid lines with
    filled circles) and the oracle approach
    (dashed lines with open squares).
    Panels (a)--(c) show the P10 recovery fraction
    as a function of the injected source-flux
    fraction $f_s$, baseline magnitude $G_0$,
    and angular Einstein radius $\theta_{\mathrm E}$,
    respectively.
    Panels (b) and (c) consider only unblended
    configurations.
    Panel (d) shows the median ratio
    $\theta_{\mathrm{E,rec}}/\theta_{\mathrm{E,inj}}$
    as a function of $f_s$.
    The shaded regions in panel (d) indicate the
    16th--84th percentile ranges, and the horizontal
    dotted line indicates perfect recovery.
    Both approaches are evaluated on the same
    configurations selected by the standard pipeline.
    }
    \label{fig:oracle_comparison}
\end{figure*}

Figure~\ref{fig:oracle_comparison} compares the
recovery of $\theta_{\mathrm E}$ using both approaches.
Although the oracle generally yields higher recovery
fractions, recovery accuracy decreases substantially
with increasing blending in both approaches
(panel a).

For unblended configurations, the recovery fraction
decreases towards fainter baseline magnitudes and
smaller Einstein radii (panels b and c).
The improvement obtained with the oracle is more
pronounced at larger Einstein radii.

Both approaches exhibit a systematic underestimation
of $\theta_{\mathrm E}$ with increasing blending,
with nearly identical median recovery ratios
(panel d).

We further compare the empirical coverage of the formal
astrometric uncertainties between the two approaches
(Table~\ref{tab:uncertainty_coverage}).
In the unblended experiment, 60.58\% and 86.21\%
of selected configurations satisfy
$|z_{\theta_{\rm E}}|\leq1$ and
$|z_{\theta_{\rm E}}|\leq2$, respectively.
For the oracle approach, the corresponding fractions
increase to 68.14\% and 95.43\%, in close agreement
with the Gaussian expectations.

As shown in Table~\ref{tab:blending_uncertainties},
coverage generally decreases with increasing blending
in both approaches. At $f_{\rm s}=0.5$, the empirical $1\sigma$
coverage decreases to 8.51\% for the standard pipeline
and 12.94\% for the oracle approach.
The corresponding median standardised residuals are
$-7.51$ and $-5.92$, respectively.

These results indicate that photometric fitting
errors contribute to the reduced accuracy of
$\theta_{\mathrm E}$ recovery, but cannot alone
account for the systematic underestimation
observed in the presence of blending.
This behaviour is consistent with the assumption
of a dark lens in the astrometric model.

\subsection{Lens mass and distance recovery}
\label{sec:MD_recovery_results}

Having examined the recovery of the angular
Einstein radius and its dependence on baseline
magnitude, blending, and photometric fitting
errors, we now investigate how accurately
the physical properties of the lens can be
recovered.

The mass--distance experiment comprises 60\,000
simulated configurations, including 20\,000
unblended configurations and 40\,000 configurations
with randomly assigned source-flux fractions,
including $f_s=1$.

Table~\ref{tab:MD_recovery} summarises the selection
and joint recovery fractions for both samples.
The selection fraction decreases from 70.25\%
in the unblended sample to 52.96\% in the
random-blending sample.

Among the selected unblended configurations,
37.65\% satisfy the joint P10 criterion and
55.20\% satisfy the joint P20 criterion.
For the random-blending sample, these fractions
decrease to 13.42\% and 27.08\%, respectively.

\begin{table}[!htbp]
\centering
    \caption{Membership selection and joint recovery
    of lens mass and distance in the mass--distance
    experiment.}
    \label{tab:MD_recovery}

    \small
    \renewcommand{\arraystretch}{1.2}
    \setlength{\tabcolsep}{3pt}

    \begin{tabular*}{\columnwidth}{
    @{\extracolsep{\fill}}lrrrrr@{}
    }
        \hline\hline
        Sample & $N_{\rm all}$ & $N_{\rm sel}$
        & Sel. & P10 & P20 \\
        & & & [\%] & [\%] & [\%] \\
        \hline
        Unblended
        & 20000 & 14050 & 70.25 & 37.65 & 55.20 \\
        Random blending
        & 40000 & 21183 & 52.96 & 13.42 & 27.08 \\
        \hline
    \end{tabular*}

    \tablefoot{
    $N_{\rm all}$ and $N_{\rm sel}$ denote the
    numbers of injected and selected configurations,
    respectively.
    Sel. denotes the selection fraction.
    P10 and P20 refer to the joint recovery of
    $M_{\mathrm L}$ and $D_{\mathrm L}$ among
    selected configurations, as defined in
    Sect.~\ref{sec:recovery_criteria}.
    The random-blending sample includes $f_s=1$.
    }
\end{table}

\begin{figure*}[!t]    \centering
    \includegraphics[width=0.9\textwidth]
    {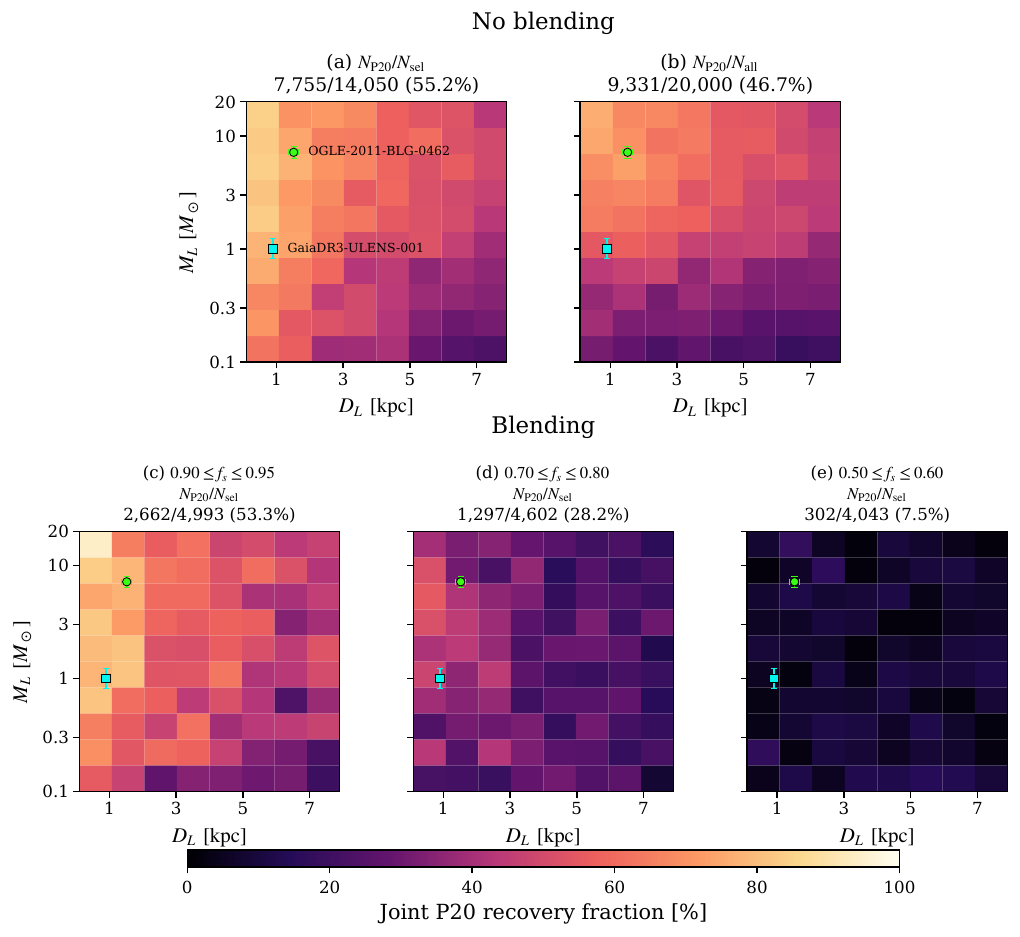}

    \caption{Joint P20 recovery fraction in the lens
mass--distance experiment, shown across
the $(M_{\mathrm L},D_{\mathrm L})$
parameter space. The upper row shows the results for the
    unblended sample, considering selected
    configurations (a) and all injected
    configurations (b).
    The lower row (c--e) shows the recovery
    fractions among selected configurations
    for different ranges of the injected
    source-flux fraction $f_s$.
    The percentages above each panel indicate
    the overall joint P20 recovery fraction
    for the corresponding sample.
    The green circles and cyan squares with
error bars indicate two observed microlensing
events: OGLE-2011-BLG-0462
($M_{\mathrm L}=7.15\pm0.83\,M_\odot$,
$D_{\mathrm L}=1.52\pm0.15$\,kpc;
\citealt{sahu2025}) and GaiaDR3-ULENS-001
($M_{\mathrm L}=1.00^{+0.23}_{-0.18}\,M_\odot$,
$D_{\mathrm L}=0.90^{+0.14}_{-0.11}$\,kpc;
\citealt{jablonska2022}), respectively.
    }
    \label{fig:mass_distance_recovery}
\end{figure*}

Figure~\ref{fig:mass_distance_recovery} shows
the joint P20 recovery fraction across the
lens mass--distance parameter space.

In the absence of blending, higher recovery
fractions are obtained for nearby and more
massive lenses.

With increasing blending, the joint recovery
fraction decreases across the mass--distance
parameter space, even among selected
configurations.

To further investigate the recovery of individual
lens properties, we consider the P20 recovery
fractions for lens mass, lens distance, and
both parameters jointly in the selected
unblended sample.

Among the 14\,050 selected unblended configurations,
65.48\% satisfy the P20 criterion for lens mass
and 69.07\% for lens distance.
Both parameters satisfy the P20 criterion in
55.20\% of configurations, while neither parameter
satisfies this criterion in 20.64\% of configurations.

\begin{figure}[t]
    \centering
    \includegraphics[width=\columnwidth]
    {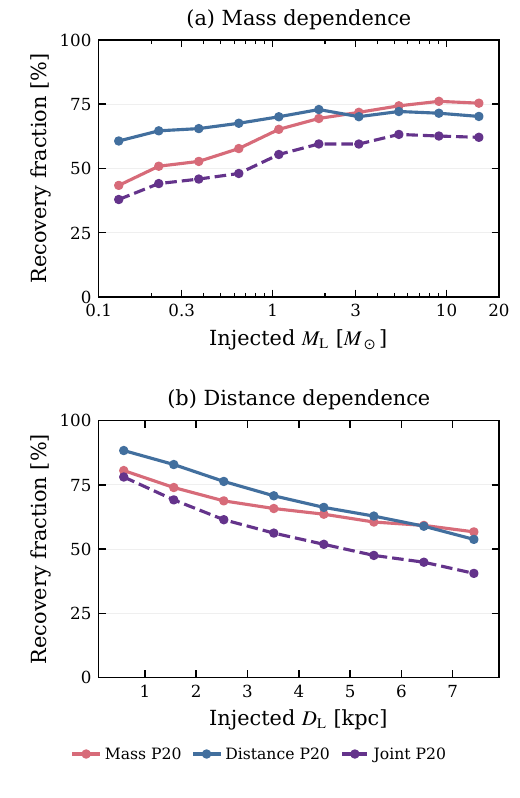}

    \caption{
    P20 recovery fractions for lens mass (pink),
    lens distance (blue), and both parameters
    jointly (dashed purple), as a function of
    the injected lens mass (a) and lens distance (b).
    Only selected configurations from the
unblended lens mass--distance sample
are included.
    Recovery fractions are calculated independently
    in each bin.
    }
    \label{fig:MD_efficiency}
\end{figure}

Figure~\ref{fig:MD_efficiency} shows these
recovery fractions as a function of the injected
lens mass and distance, aggregated over the
remaining simulated parameters.

For low-mass lenses, the recovery fraction of
the lens distance is higher than that of the
lens mass.
This difference decreases towards larger lens
masses.
The recovery fractions of both parameters
decrease with increasing lens distance.
The joint recovery fraction follows the same
general trends.

Finally, we examine the occurrence of non-positive
parameter estimates.
Among the 35\,233 selected configurations in
the mass--distance experiment, 1\,661 (4.71\%)
yield negative angular Einstein-radius estimates,
corresponding to negative inferred lens masses,
1\,027 (2.91\%) yield negative inferred lens
parallaxes, and 1\,956 (5.55\%) yield non-positive
source-parallax estimates.
These categories are not mutually exclusive.

Overall, 3\,457 of the 35\,233 selected
configurations (9.81\%) fail at least one
positivity condition. The fractions are
similar in the unblended (9.91\%) and
random-blending (9.75\%) samples.
Of these 3\,457 configurations, 77 satisfy
the joint P10 criterion and 195 satisfy
the joint P20 criterion.

\section{Discussion}
\label{sec:discussion}
\subsection{Parameter recovery accuracy}
\label{sec:discussion_recovery}

Our results demonstrate that the angular Einstein
radius is recovered more accurately for brighter
baseline magnitudes and larger injected
$\theta_{\mathrm E}$, particularly in the absence
of blending
(Figs.~\ref{fig:blending_grid_p10}
and~\ref{fig:oracle_comparison}).
This is consistent with the expected signal-to-noise
dependence of astrometric microlensing, since larger
Einstein radii produce larger astrometric
displacements, while fainter objects are associated
with larger measurement uncertainties.
Similar trends were reported by \citet{jankovic2025}
for astrometry-only parameter recovery.

The recovery of individual photometric and astrometric
parameters is illustrated in
Fig.~\ref{fig:member_recovery}.
While $G_0$, $t_0$, and the source proper motions
are generally well recovered, larger scatter is
observed for the microlensing parallax components
and the source parallax.
The recovery of these parameters is particularly
important for reliable lens mass and distance
determination.

Even in the absence of blending, the recovered
values of $t_{\mathrm E}$ and $\theta_{\mathrm E}$
exhibit asymmetric scatter towards lower values.
The underestimation of $t_{\rm E}$ may be related
to degeneracies between photometric parameters
and limitations of the observational sampling.
Inaccuracies in the photometric solution may
subsequently affect the recovery of
$\theta_{\mathrm E}$ during astrometric fitting.

The close agreement between the empirical uncertainty
coverage and Gaussian expectations in the unblended
oracle experiment suggests that errors in the fixed
photometric parameters contribute to the lower
coverage obtained with the standard pipeline.
Nevertheless, the increase in recovery accuracy with
$S_{\theta_{\rm E}}$ indicates that formal measurement
significance provides a useful additional diagnostic
of solution reliability in the absence of blending.

In some configurations, the recovered value of $u_0$
has the opposite sign to its injected value. The two photometric solutions may be difficult
to distinguish, particularly when the microlensing
parallax signal is weak.
In our sequential approach, astrometric information
provides an additional constraint on the choice
between the two photometric branches.
However, this does not always resolve the sign
degeneracy.

The limited observational baseline and irregular
sampling of Gaia introduce further limitations
on parameter recovery.
Short events may be poorly sampled, while longer
events may extend beyond the Gaia observing period
\citep{jankovic2025}.
Part of the astrometric microlensing signal may
also be absorbed by the standard five-parameter
astrometric model \citep{jablonska2022,kaczmarek2026b}.

The relatively small differences in recovery
fractions between the three investigated sky
positions (Table~\ref{tab:sky_comparison})
suggest that the number of transits alone is
not sufficient to characterise astrometric
recovery. The distribution of observation epochs
and scanning angles relative to the microlensing
trajectory should also be considered.

\subsection{Impact of blending}
\label{sec:discussion_blending}

Blending constitutes a major limitation of the
sequential fitting approach investigated in this
study.
The residual distributions presented in
Appendix~\ref{app:residuals} reveal systematic
underestimation of $t_{\mathrm E}$ and
$\theta_{\mathrm E}$ with increasing blending.

As follows from Eq.~\eqref{eq:photometric_model},
blending reduces the observed photometric
magnification. When blending is neglected,
this reduction may be compensated for by
larger $|u_0|$ and shorter $t_{\mathrm E}$
\citep{wozniak1997}.
This degeneracy may explain the observed
underestimation of $t_{\mathrm E}$.

In the astrometric fit, neglecting the
contribution of lens light to the photocentre
motion may lead to a reduction of the fitted
$\theta_{\mathrm E}$, which scales the
astrometric microlensing trajectory.

However, photometric fitting errors are not
the only source of systematic bias.
The oracle experiment
(Fig.~\ref{fig:oracle_comparison}d)
demonstrates that $\theta_{\mathrm E}$ remains
systematically underestimated even when the
injected photometric parameters are used.
The nearly identical median recovery ratios
obtained with both approaches highlight
the importance of accounting for lens light
in the astrometric model \citep{han1999}.

The standardised residuals and empirical uncertainty
coverage (Table~\ref{tab:blending_uncertainties})
further demonstrate that the systematic
underestimation of $\theta_{\mathrm E}$ is not
adequately reflected in its formal uncertainty.
The low coverage obtained even with the oracle
approach indicates that this discrepancy cannot
be attributed solely to errors in the recovered
photometric parameters.
Consequently, even solutions with high
$\theta_{\mathrm{E,rec}}/\sigma_{\theta_{\mathrm E}}$
can exhibit substantial systematic errors.
Formal measurement precision should therefore
not be interpreted as a guarantee of recovery
accuracy in the presence of blending.

Blending also affects the recovery of the
source parallax. As discussed by \citet{jankovic2025},
in the absence of magnification, the parallax of
the unresolved source--lens system can be expressed as
\begin{equation}
    \varpi_{\mathrm G}
    = f_s\varpi_{\mathrm S}
    + (1-f_s)\varpi_{\mathrm L}.
\end{equation}
Since the lens is located in front of the source,
$\varpi_{\mathrm L}>\varpi_{\mathrm S}$, and
therefore the photocentre parallax is larger
than the source parallax for $f_s<1$.
This effect may explain the systematic
overestimation of $\varpi_{\mathrm S}$
observed in Appendix~\ref{app:residuals}.

These results highlight the limitations of the
dark-lens approximation when interpreting
microlensing parameters in the presence of blending.

\subsection{Constraining the lens properties}
\label{sec:discussion_lens_properties}

The higher recovery fractions for nearby and more
massive lenses can be partly attributed to larger
angular Einstein radii and hence larger astrometric
microlensing signals.
This trend is consistent with the findings of
\citet{jankovic2025}, who reported a recovery
fraction of approximately 23\% under their
20\% accuracy criterion.
In our unblended sample, the joint P20 recovery
fraction reaches 46.66\% among all injected
configurations (Fig.~\ref{fig:mass_distance_recovery}b).
However, differences in parameter sampling,
recovery criteria, and fitting procedures
limit a direct quantitative comparison.

The recovery fractions increase with lens mass
but flatten at higher masses
(Fig.~\ref{fig:MD_efficiency}a).
At fixed lens and source distances, a higher lens
mass leads to a larger $\theta_{\rm E}$ but a smaller
$\pi_{\rm E}$.
Thus, despite a stronger astrometric signal,
the microlensing parallax may be more difficult
to measure.
A longer observing period may improve parameter
recovery, particularly for long-timescale events.

The recovery fractions provide a reference
for identifying the regions of the lens
mass--distance parameter space in which the
sequential approach performs most reliably.
These results are particularly relevant to
searches for dark compact objects, including
isolated stellar-mass black holes.
The systematic underestimation of
$\theta_{\mathrm E}$ in blended configurations
may lead to underestimated lens masses when
the microlensing parallax is accurately determined.
However, uncertainties in $\pi_{\mathrm E}$
prevent the recovered masses from being interpreted
as strict lower limits \citep{kaczmarek2026}. Furthermore, the limited empirical coverage of
the formal $\theta_{\mathrm E}$ uncertainties
highlights the need for caution when estimating
uncertainties in the derived lens properties.

\subsection{Reliability of the membership criterion}
\label{sec:discussion_membership}
The photometric membership score is a heuristic
criterion designed for microlensing candidate
selection, not a probability of accurate
parameter recovery.

In the unblended experiment
(Table~\ref{tab:unblended_recovery}),
75.75\% of selected configurations recover
$\theta_{\mathrm E}$ within 10\% of its true value,
compared with 44.97\% of successfully fitted
configurations rejected by the membership criterion.
Thus, membership selection favours accurately
recovered solutions but also excludes some of them.

Conversely, a high membership score does not
guarantee accurate astrometric recovery,
particularly in the presence of blending.
At $f_s=0.5$, 38.01\% of configurations pass
the membership criterion, yet only 4.59\%
of these selected configurations recover
$\theta_{\mathrm E}$ within 10\%
(Table~\ref{tab:blending_recovery}).

A similar limitation is seen in the
mass--distance experiment
(Table~\ref{tab:MD_recovery}).
Although blending substantially reduces the joint
recovery fraction, the fractions of configurations
with non-positive parameter estimates remain
similar in both samples.
Negative parallax estimates may arise from
measurement noise and do not necessarily indicate
an unsuccessful astrometric fit. However, such
estimates cannot be directly interpreted as
positive physical distances through inverse parallax.
Thus, numerical recovery accuracy and the physical
interpretability of parameter estimates are
distinct aspects of solution reliability.

Unlike the photometric datasets for which the
membership score was designed, all simulated
configurations in this study represent microlensing
events, including parameter combinations that may
be rare in real events.
Penalising extreme parameter values, small
photometric amplitudes, and large parameter
uncertainties is an intentional part of the
original selection procedure.
However, extreme parameter values do not necessarily
imply poor parameter recovery.
When applied to known microlensing events,
these criteria may reject accurately recovered
configurations and introduce selection biases
that should be considered in future microlensing
population studies.

Future work could explore additional selection
criteria incorporating astrometric information.
The formal significance $S_{\theta_{\mathrm E}}$
provides a potential diagnostic of recovery
accuracy, although its reliability is limited
in the presence of blending. A ~more detailed treatment of blending, particularly in astrometric modelling, would be a valuable direction for future work, given the complexity of real Gaia observations.

\subsection{Methodological limitations and future improvements}
\label{sec:discussion_limitations}

The simulated parameter distributions were designed
to explore a broad range of microlensing configurations
rather than to represent a realistic Galactic population. In particular, the independent sampling of $\theta_{\rm E}$
and $t_{\rm E}$ in the unblended and blending experiments
does not preserve a realistic distribution of relative
proper motions, $\mu_{\rm rel}=\theta_{\rm E}/t_{\rm E}$.
Given the strong dependence of recovery accuracy on
$\theta_{\rm E}$, the overall recovery fractions are
sensitive to the adopted input distribution.
Furthermore, the limited number of investigated sky
positions does not capture the full diversity of Gaia
scanning coverage. The reported recovery fractions
should therefore not be interpreted as expected
recovery rates for the entire Gaia DR4 microlensing
population.

Some simulated configurations share the same
underlying events and are therefore not
statistically independent. The reported recovery
fractions describe the simulated sample, but
this dependence should be taken into account
when estimating confidence intervals.

Despite the use of multiple starting points in the
photometric optimisation, convergence to the global minimum is not guaranteed. The empirical scaling of photometric uncertainties
by a factor of $\sqrt{10}$ reduces the photometric
$\chi^2$ values by a factor of ten for fixed
residuals. It also increases the effective
threshold of the initial photometric selection
from $3\sigma_{\mathrm{tr}}$ to
$3\sqrt{10}\,\sigma_{\mathrm{tr}}$.
Furthermore, the scaling affects the estimated
photometric parameter uncertainties and may
influence the membership scores.

The reported recovery fractions are therefore
conditional on the adopted uncertainty scaling.
This assumption, together with differences in
the implementation of the Nelder--Mead optimiser,
convergence criteria, and optimisation limits,
restricts how closely our Python implementation
reproduces the original Gaia photometric pipeline.
Photometric parameter uncertainties were estimated from
one-dimensional sections of the $\chi^2$ surface,
neglecting parameter correlations.
The formal astrometric uncertainties are conditional
on the fixed photometric parameters and do not account
for their uncertainties.
The empirical coverage of the formal uncertainties in $\theta_{\rm E}$ is consistent
with the Gaussian expectations in the unblended oracle
experiment but is lower in the standard sequential
approach and in the presence of blending.
The calibration of uncertainties for the remaining
parameters and the derived lens properties was not
investigated.
Simultaneous photometric and astrometric modelling
could improve the propagation of parameter uncertainties
\citep{kaczmarek2026}.

A more detailed treatment of blending, particularly
in astrometric modelling, would be a valuable direction
for future work, given the complexity of real Gaia
observations.

\section{Conclusions}

We investigated the performance of sequential
photometric and astrometric modelling of
microlensing events using simulated Gaia DR4
observations. Our main conclusions are as follows:

\begin{itemize}

\item In the absence of blending, recovery accuracy
increases towards brighter baseline magnitudes
and larger angular Einstein radii.
Among selected configurations, the P10 recovery
fraction increases from 33.52\% for injected
$\theta_{\mathrm E}<1$ mas to 86.72\% for
$\theta_{\mathrm E}\geq6$ mas, with corresponding
P20 fractions of 49.11\% and 92.58\%.

\item Despite substantial differences in the
number of Gaia transits, the three investigated
sky positions yield similar recovery fractions.
This suggests that the number of transits alone
is not sufficient to characterise the accuracy
of astrometric parameter recovery.

\item Blending significantly reduces parameter
recovery accuracy, leading to a systematic
underestimation of the Einstein timescale and
angular Einstein radius, and a tendency to
overestimate the source parallax.
The systematic underestimation of
$\theta_{\mathrm E}$ persists even when the
injected photometric parameters are used,
highlighting the limitations of the dark-lens
approximation.

\item Formal measurement significance provides
a useful diagnostic of recovery accuracy in
the absence of blending.
Among selected unblended configurations with
$S_{\theta_{\mathrm E}}\geq10$, 91.38\%
satisfy P10 and 96.81\% satisfy P20.
However, high formal significance does not
guarantee accurate recovery in the presence
of blending.

\item The empirical coverage of the formal
$\theta_{\mathrm E}$ uncertainties is consistent
with Gaussian expectations in the unblended
oracle experiment but is lower in the standard
pipeline. In the presence of blending, both
approaches yield substantially reduced coverage,
indicating that the formal uncertainties do not
adequately account for systematic recovery errors.

\item In the absence of blending, the lens mass
and distance are recovered within 20\% of their
injected values for 65.48\% and 69.07\% of selected
configurations, respectively, while 55.20\%
satisfy both criteria simultaneously.
The joint recovery fraction decreases to
27.08\% in the random-blending sample.
The highest joint recovery fractions are
obtained for nearby and massive lenses.

\item The photometric membership criterion favours
accurate parameter recovery but does not guarantee
it and may also reject accurately recovered
configurations. Additional selection criteria
incorporating astrometric diagnostics could
improve the reliability of recovered parameter
estimates.

\end{itemize}

The recovery fractions obtained in this study
characterise the performance of our sequential
fitting approach under the adopted simulation
conditions. Given the adopted parameter
distributions, simplified modelling
assumptions, limited sky coverage, and differences
from the original Gaia pipeline, these fractions
should not be interpreted as predictions for
the full Gaia DR4 microlensing population.
Nevertheless, our results provide a useful
reference for future microlensing investigations
with Gaia DR4.
The longer observational baseline of Gaia DR5
may further improve parameter recovery for
long-timescale events and reduce the limitations
observed at higher lens masses
(Fig.~\ref{fig:MD_efficiency}a).

With the large volumes of data from Gaia,
the Vera C. Rubin Observatory's Legacy Survey
of Space and Time (LSST), and the Nancy Grace
Roman Space Telescope, computationally efficient
methods for selecting and characterising
microlensing events will become increasingly
important \citep{jankovic2025,abrams2025,kaczmarek2025}.
The sequential approach offers a practical
alternative to simultaneous photometric and
astrometric fitting.
The selection criteria involve a trade-off between
the number of retained events and the accuracy
of the recovered parameters: some accurately
recovered configurations may be rejected to
reduce the fraction of unreliable solutions
in the selected sample.

Overall, our results demonstrate that the
sequential approach can accurately recover
microlensing parameters under favourable
observational conditions.
Despite the limitations identified in this work,
the pipeline provides a practical way to narrow
down a large sample of microlensing events to
a subset with a higher fraction of accurately
recovered parameters, suitable for further
investigation.

\
\begin{acknowledgements}
{\L}.W. and M.W. acknowledge support from the Polish National Science Centre DAINA grant No 2024/52/L/ST9/00210.
{\L}.W. acknowledge support from the European Union's Horizon Europe Research and Innovation programme ACME under grant agreement No 101131928 (2024-2028). 
This work has made use of data from the European Space Agency (ESA) mission \textit{Gaia} (\url{https://www.cosmos.esa.int/gaia}), processed by the \textit{Gaia} Data Processing and Analysis Consortium (DPAC; \url{https://www.cosmos.esa.int/web/gaia/dpac/consortium}). Funding for DPAC has been provided by national institutions, in particular the institutions participating in the \textit{Gaia} Multilateral Agreement.
\end{acknowledgements}


\bibliographystyle{bibtex/aa}
\bibliography{references}

\begin{appendix}

\section{Photometric and astrometric parameter recovery}
\label{app:parameter_recovery}

\begin{figure*}[p]
    \centering
    \includegraphics[
        width=0.95\textwidth
    ]{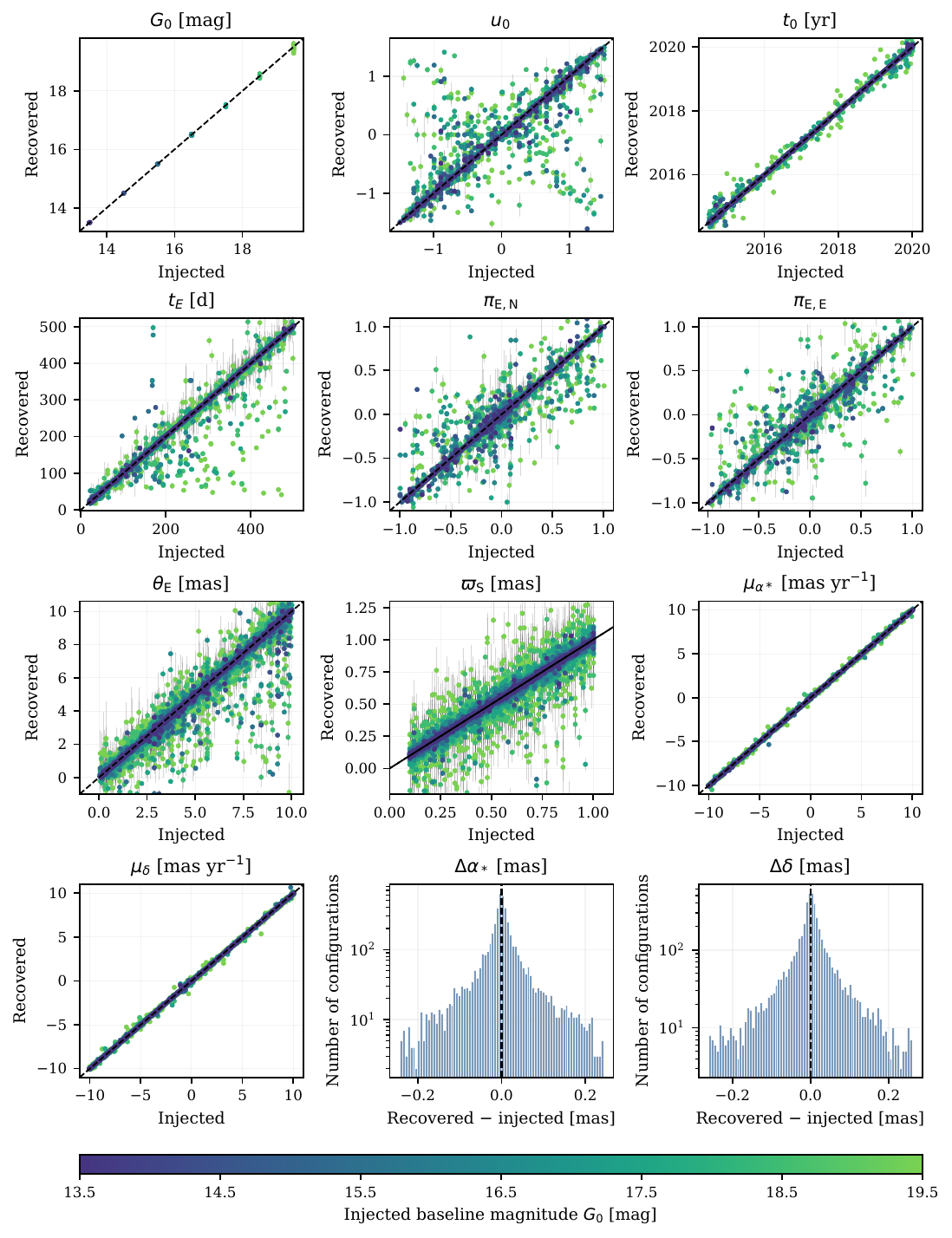}

    \caption{
    Comparison of injected and recovered photometric
    (upper six panels) and astrometric (lower six
    panels) parameters for 5\,000 randomly drawn
    unblended configurations with successful
    astrometric fits and photometric membership
    scores $\geq0.8$.
    Dashed lines indicate perfect recovery, colours
    represent the injected baseline magnitude $G_0$,
    and error bars indicate the estimated parameter
    uncertainties.
    The two lower-right panels show the distributions
    of source-position residuals with logarithmic
    number counts.
    }
    \label{fig:member_recovery}
\end{figure*}

Figure~\ref{fig:member_recovery} compares the injected
and recovered photometric and astrometric parameters
for 5\,000 randomly drawn configurations from the
selected unblended sample described in
Sect.~\ref{sec:results_unblended}.

The baseline magnitude $G_0$, reference epoch $t_0$,
and source proper motions $\mu_{\alpha*,\mathrm S}$
and $\mu_{\delta,\mathrm S}$ show relatively small
scatter. Larger deviations are observed for $u_0$,
$t_{\mathrm E}$, the microlensing parallax
components $\pi_{\mathrm{E,N}}$ and
$\pi_{\mathrm{E,E}}$, the angular Einstein radius
$\theta_{\mathrm E}$, and the source parallax
$\varpi_{\mathrm S}$.

In some configurations, $u_0$ is recovered with
the opposite sign, reflecting the degeneracy
between the positive and negative $u_0$ solutions.
The distributions of the source-position parameter
residuals peak near zero, with extended tails
on both sides.

Parameter uncertainties were estimated as described
in Sects.~\ref{sec:photometric_optimisation}
and~\ref{sec:linear_astrometric_fitting}.

\section{Dependence on Gaia scanning coverage}
\label{app:additional_results}

Table~\ref{tab:sky_comparison} presents the detailed
selection and angular Einstein-radius recovery fractions
for the three investigated sky positions, corresponding
to different Gaia scanning coverage.

\begin{table}[!htbp]
\centering
\caption{Membership selection and angular Einstein-radius
recovery at three sky positions in the unblended
sample ($f_s=1$).}
\label{tab:sky_comparison}

\small
\renewcommand{\arraystretch}{1.2}
\setlength{\tabcolsep}{5pt}

\begin{tabular*}{\columnwidth}{
@{\extracolsep{\fill}}lcccc@{}
}
\hline\hline
Criterion & Field 1 & Field 2 & Field 3
& \textbf{Total} \\
\hline
FoV transits
    & 268 & 216 & 94 & -- \\
$N_{\rm all}$
    & 9786 & 9786 & 9786 & \textbf{29358} \\
$N_{\rm sel}$
    & 5042 & 4900 & 4905 & \textbf{14847} \\
Sel. [\%]
    & 51.52 & 50.07 & 50.12 & \textbf{50.57} \\
\hline
P10 [\%]
    & 77.09 & 76.92 & 73.19 & \textbf{75.75} \\
P20 [\%]
    & 86.30 & 86.24 & 83.36 & \textbf{85.31} \\
\hline
\end{tabular*}

\tablefoot{
Fields 1, 2, and 3 correspond to
$(\alpha,\delta)=(6.5^\circ,-47.3^\circ)$,
$(277.8^\circ,-68.3^\circ)$, and
$(287.9^\circ,45.9^\circ)$, respectively.
$N_{\rm all}$ and $N_{\rm sel}$ denote the numbers
of injected and selected configurations, respectively.
Sel. denotes the selection fraction.
P10 and P20 are calculated among selected
configurations, as defined in
Sect.~\ref{sec:recovery_criteria}.
}
\end{table}

\section{Selection and recovery fractions
in the blending experiment}
\label{app:blending_recovery}

Table~\ref{tab:blending_recovery} presents
the selection fractions and angular Einstein-radius
recovery fractions for all eleven injected
source-flux fractions considered in
Sect.~\ref{sec:blending_recovery}.

\begin{table}[!htbp]
\centering
\caption{Membership selection and angular Einstein-radius
recovery in the blending experiment.}
\label{tab:blending_recovery}

\small
\renewcommand{\arraystretch}{1.2}
\setlength{\tabcolsep}{3pt}

\begin{tabular*}{\columnwidth}{
@{\extracolsep{\fill}}rrrrrr@{}
}
\hline\hline
$f_s$ & $N_{\rm all}$ & $N_{\rm sel}$
& Sel. & P10 & P20 \\
& & & [\%] & [\%] & [\%] \\
\hline
0.10 & 10206 & 1135 & 11.12 & 0.62 & 1.06 \\
0.20 & 10206 & 2132 & 20.89 & 0.66 & 1.22 \\
0.30 & 10206 & 2884 & 28.26 & 1.46 & 3.02 \\
0.40 & 10206 & 3467 & 33.97 & 2.37 & 7.59 \\
0.50 & 10206 & 3879 & 38.01 & 4.59 & 25.65 \\
0.60 & 10206 & 4335 & 42.48 & 15.39 & 57.42 \\
0.70 & 10206 & 4636 & 45.42 & 37.86 & 75.45 \\
0.80 & 10206 & 4939 & 48.39 & 63.68 & 82.73 \\
0.90 & 10206 & 5162 & 50.58 & 74.85 & 85.63 \\
0.95 & 10206 & 5195 & 50.90 & 76.30 & 86.08 \\
1.00 & 10206 & 5201 & 50.96 & 77.16 & 86.29 \\
\hline
\textbf{Total} &
\textbf{112266} &
\textbf{42965} &
\textbf{38.27} &
\textbf{41.27} &
\textbf{57.81} \\
\hline
\end{tabular*}

\tablefoot{
$N_{\rm all}$ and $N_{\rm sel}$ denote the numbers
of injected and selected configurations, respectively.
Sel. denotes the selection fraction.
P10 and P20 are calculated among selected
configurations, as defined in
Sect.~\ref{sec:recovery_criteria}.
The total refers to all source-flux fractions combined.
}
\end{table}

\section{Parameter recovery residuals}
\label{app:residuals}

\begin{figure*}[t]
    \centering
    \includegraphics[width=\textwidth]
    {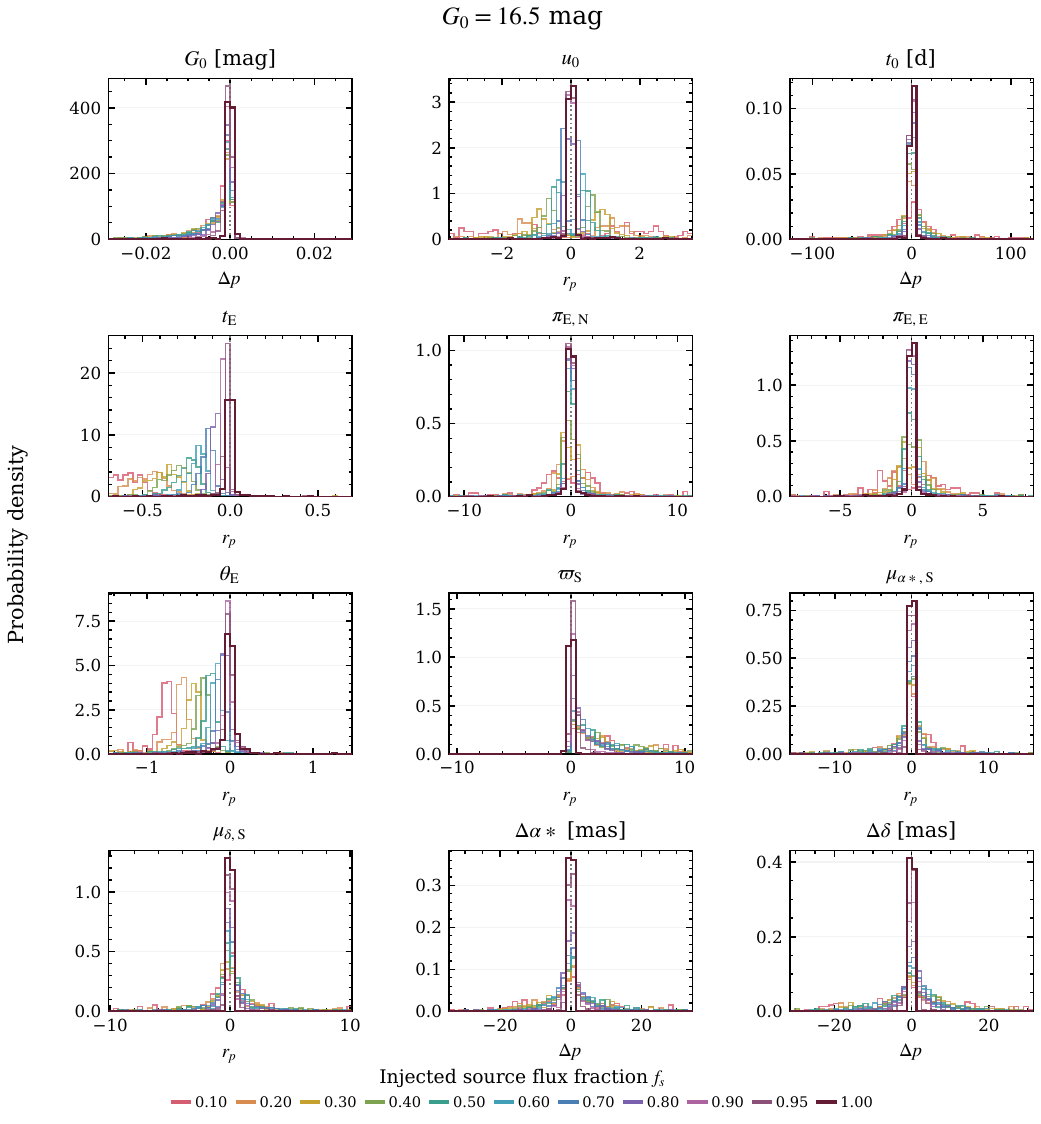}

    \caption{
    Residual distributions of photometric and
    astrometric parameters for selected configurations
    at an injected baseline magnitude of
    $G_0=16.5$ mag.
    Colours indicate the injected source-flux
    fractions $f_s$.
    Signed residuals are defined as
    $\Delta p=p_{\mathrm{rec}}-p_{\mathrm{inj}}$,
    and relative residuals as
    $r_p=(p_{\mathrm{rec}}-p_{\mathrm{inj}})
    /|p_{\mathrm{inj}}|$.
    Relative residuals are excluded when
    $p_{\mathrm{inj}}=0$. 
Signed residuals are shown for
$G_0$, $t_0$, $\Delta\alpha^*$, and
$\Delta\delta$, while the remaining
panels show relative residuals.
    The vertical dotted lines indicate perfect
    recovery.
    The horizontal ranges are limited to the
    98th percentile of the absolute residuals
    across all baseline magnitudes and source-flux
    fractions in the selected sample.
    Each distribution is normalised using all
    finite residuals in the corresponding group,
    including those outside the displayed range.
    }
    \label{fig:residual_distributions}
\end{figure*}

Figure~\ref{fig:residual_distributions} shows
the residual distributions of the recovered
photometric and astrometric parameters for
selected configurations from the blending
experiment at $G_0=16.5$ mag.

Increasing blending generally leads to broader
residual distributions. Systematic underestimation
is observed for the Einstein timescale
$t_{\mathrm E}$ and angular Einstein radius
$\theta_{\mathrm E}$, while the source parallax
$\varpi_{\mathrm S}$ tends to be overestimated.

The systematic underestimation of
$\theta_{\mathrm E}$ is also observed in the
oracle experiment (Sect.~\ref{sec:oracle_recovery}),
where the injected photometric parameters
are used during astrometric fitting.

\clearpage
\onecolumn

\section{Einstein-radius recovery and formal uncertainties}
\label{app:uncertainties}

Table~\ref{tab:theta_snr_recovery} presents the P10 and P20
recovery fractions as a function of the injected angular
Einstein radius and the formal significance of the recovered
value in the unblended experiment.
The empirical coverage of the formal uncertainties is
summarised in Table~\ref{tab:uncertainty_coverage}.
The dependence on the injected source-flux fraction is
presented in Table~\ref{tab:blending_uncertainties}.


\noindent
\begin{minipage}[t]{0.48\textwidth}
\vspace{0pt}

\begin{table}[H]
\centering
\caption{Angular Einstein-radius recovery as a function of
the injected Einstein radius and formal measurement
significance in the unblended experiment.}
\label{tab:theta_snr_recovery}
\small
\setlength{\tabcolsep}{3pt}

\begin{tabular}{lrrrr}
\hline\hline
Interval & $N_{\rm all}$ & $N_{\rm sel}$ & P10 & P20 \\
         &               &               & [\%] & [\%] \\
\hline
\multicolumn{5}{l}{Injected $\theta_{\rm E}$ [mas]} \\
$[0.05,1)$ & 2499  & 1238 & 33.52 & 49.11 \\
$[1,3)$    & 5649  & 2923 & 66.51 & 81.05 \\
$[3,6)$    & 9366  & 4746 & 78.72 & 88.28 \\
$[6,10]$   & 11844 & 5940 & 86.72 & 92.58 \\
\hline
\multicolumn{5}{l}{Formal significance $S_{\theta_{\rm E}}$} \\
$S<0$          & -- & 191   & 0.00  & 0.00  \\
$0\leq S<3$    & -- & 1050  & 11.52 & 23.52 \\
$3\leq S<5$    & -- & 823   & 28.19 & 49.70 \\
$5\leq S<10$   & -- & 1842  & 48.59 & 76.98 \\
$S\geq10$      & -- & 10941 & 91.38 & 96.81 \\
\hline
\end{tabular}

\tablefoot{
$N_{\rm all}$ and $N_{\rm sel}$ denote the numbers of
injected and selected configurations, respectively.
P10 and P20 are calculated among selected configurations
in each interval.
The formal significance is defined as
$S_{\theta_{\rm E}} =
\theta_{\rm E,rec}/\sigma_{\theta_{\rm E}}$.
The formal-significance analysis is restricted to the
selected sample.
}
\end{table}

\end{minipage}%
\hfill
\begin{minipage}[t]{0.48\textwidth}
\vspace{0pt}


\begin{table}[H]
\centering
\caption{Empirical coverage of the formal uncertainties
of the angular Einstein radius.}
\label{tab:uncertainty_coverage}
\small
\setlength{\tabcolsep}{3pt}

\begin{tabular}{lrrrr}
\hline\hline
Sample & Median $z$ & $C_1$ & $C_2$ & $C_3$ \\
       &            & [\%] & [\%] & [\%] \\
\hline
Gaussian reference & 0.00 & 68.27 & 95.45 & 99.73 \\
\hline
\multicolumn{5}{l}{Unblended experiment ($N_{\rm sel}=14847$)} \\
Standard pipeline & $-0.09$ & 60.58 & 86.21 & 92.34 \\
Oracle            & $ 0.01$ & 68.14 & 95.43 & 99.72 \\
\hline
\multicolumn{5}{l}{Blending experiment ($N_{\rm sel}=42965$)} \\
Standard pipeline & $-2.67$ & 26.52 & 42.78 & 51.54 \\
Oracle            & $-1.82$ & 30.36 & 48.46 & 57.28 \\
\hline
\end{tabular}

\tablefoot{
The standardised residual is defined as
$z=(\theta_{\rm E,rec}-\theta_{\rm E,inj})/
\sigma_{\theta_{\rm E}}$.
$C_k$ denotes the percentage of configurations satisfying
$|z|\leq k$.
Both approaches are evaluated on the same configurations
selected by the standard pipeline.
The blending experiment includes all investigated
source-flux fractions, including $f_{\rm s}=1$.
}
\end{table}

\end{minipage}

\par
\vspace{0.5cm}


\begin{table}[H]
\centering
\caption{Formal measurement significance, angular
Einstein-radius recovery, and empirical uncertainty
coverage as a function of the injected source-flux fraction.}
\label{tab:blending_uncertainties}

\begin{tabular}{rrrrrrrrrrrr}
\hline\hline
& & \multicolumn{4}{c}{$S_{\theta_{\rm E}}\geq10$}
& \multicolumn{3}{c}{Standard pipeline}
& \multicolumn{3}{c}{Oracle} \\
\cline{3-6}\cline{7-9}\cline{10-12}
$f_{\rm s}$ & $N_{\rm sel}$ & $N_{S\geq10}$ & Fraction
& P10 & P20
& Median $z$ & $C_1$ & $C_2$
& Median $z$ & $C_1$ & $C_2$ \\
& & & [\%] & [\%] & [\%]
& & [\%] & [\%] & & [\%] & [\%] \\
\hline
0.10 & 1135 & 337  & 29.69 & 0.30  & 0.89
& -21.05 & 3.35 & 6.70
& -21.62 & 1.85 & 5.02 \\
0.20 & 2132 & 991  & 46.48 & 0.00  & 0.10
& -16.62 & 3.33 & 7.60
& -15.60 & 3.52 & 8.16 \\
0.30 & 2884 & 1599 & 55.44 & 0.38  & 0.69
& -12.81 & 4.96 & 9.78
& -11.27 & 5.69 & 11.55 \\
0.40 & 3467 & 2145 & 61.87 & 0.75  & 3.68
& -10.09 & 6.23 & 13.09
& -8.22 & 9.32 & 17.62 \\
0.50 & 3879 & 2569 & 66.23 & 2.37  & 26.78
& -7.51 & 8.51 & 17.12
& -5.92 & 12.94 & 24.44 \\
0.60 & 4335 & 3025 & 69.78 & 15.27 & 67.11
& -5.28 & 12.78 & 25.10
& -3.73 & 18.39 & 33.15 \\
0.70 & 4636 & 3358 & 72.43 & 43.27 & 87.70
& -3.53 & 19.07 & 33.84
& -2.37 & 23.30 & 40.83 \\
0.80 & 4939 & 3650 & 73.90 & 76.05 & 93.95
& -2.10 & 27.78 & 47.42
& -1.35 & 31.30 & 52.50 \\
0.90 & 5162 & 3860 & 74.78 & 89.09 & 96.30
& -1.04 & 40.45 & 64.26
& -0.63 & 43.37 & 69.20 \\
0.95 & 5195 & 3908 & 75.23 & 90.63 & 96.44
& -0.60 & 49.59 & 76.44
& -0.32 & 53.07 & 81.79 \\
1.00 & 5201 & 3934 & 75.64 & 91.31 & 96.77
& -0.10 & 60.01 & 85.68
& 0.00 & 68.06 & 95.25 \\
\hline
\end{tabular}

\tablefoot{
$N_{\rm sel}$ denotes the number of configurations selected
by the standard pipeline.
$N_{S\geq10}$ is the number of selected configurations
satisfying $S_{\theta_{\rm E}}\geq10$ in the same pipeline.
Fraction denotes their percentage relative to $N_{\rm sel}$.
P10 and P20 are calculated exclusively among configurations
satisfying this significance threshold.
In contrast, the median standardised residuals and empirical
coverage fractions $C_1$ and $C_2$ are calculated over all
selected configurations, irrespective of their formal
significance, for both the standard pipeline and the oracle.
}
\end{table}





\end{appendix}
\end{document}